# Luminosity Measurement at ATLAS – Development, Construction and Test of Scintillating Fibre Prototype Detectors

S. Ask[a], P. Barrillon[b], A. Braem[a], C. Cheiklali[b], I. Efthymiopoulos[a], D. Fournier[b], C. de La Taille[b], B. Di Girolamo[a], P. Grafstrom[a], C. Joram[a*], M. Haguenauer[c], V. Hedberg[d] B. Lavigne[b], A. Maio[e], A. Mapelli[a], U. Mjörnmark[d], P. Puzo[b], M. Rijssenbeek[f], J. Santos[e], J.G. Saraiva[e], H. Stenzel[g], M. Thioye[f], E. Valladolid[a] and V. Vorobel[h]

[a]CERN, PH Department, Geneva, Switzerland

[b]Laboratoire d'Accelerateur Lineaire, Orsay, France

[c]Ecole Polytechnique, Palaiseau, France

[d]University of Lund, Sweden

[e]LIP – Laboratório de Instrumentação e Física Experimental de Partículas and CFNUL – Centro de Física Nuclear da Universidade de Lisboa, Lisbon, Portugal

[f]Stony Brook University, New York, USA

[g]II. Physikalisches Institut, Justus-Liebig-Universität, Giessen, Germany

[h]Faculty of Mathematics and Physics, Charles University in Prague, Czech Republic

ABSTRACT

We are reporting about a scintillating fibre tracking detector which is proposed for the precise determination of the absolute luminosity of the CERN LHC at interaction point 1 where the ATLAS experiment is located. The detector needs to track protons elastically scattered under $\mu$rad angles in direct vicinity to the LHC beam. It is based on square shaped scintillating plastic fibres read out by multi-anode photomultiplier tubes and is housed in Roman Pots. We describe the design and construction of prototype detectors and the results of a beam test experiment at DESY. The excellent detector performance established in this test validates the detector design and supports the feasibility of the proposed challenging method of luminosity measurement.

## 1. Introduction

A system of ultra-small-angle detectors, located at 240 m on either side of the main ATLAS detector at CERN, is proposed to measure elastically scattered protons for the primary purpose of absolute determination of the LHC luminosity at the ATLAS interaction point (IP) [1]. The system is called ALFA (Absolute Luminosity For ATLAS) and consists of *Roman Pot* inserts in the beam pipe, equipped with position sensitive detectors. With appropriate beam optics it will allow detection of scattered protons at small enough ($\approx$ 1 mm) distances away from the circulating beam, to reach the theoretically well-calculable Coulomb scattering regime. ALFA is complemented by the relative luminosity monitor LUCID (LUminosity measurement using a Cerenkov Integrating Detector) [1] which is based on cylindrical Cerenkov counters placed around the beam pipe close to the interaction point. LU-

*Corresponding author, Christian.Joram@cern.ch

CID will thus be employed to monitor the instantaneous luminosity for a given period and will be calibrated by ALFA via Coulomb scattering in dedicated luminosity runs. The luminosity calibration will be obtained from the measurement of the $t$-spectrum[2] down to scattering angles of 2.7 $\mu$rad covering the rise of the spectrum induced by Coulomb interaction, from which the absolute luminosity is extracted independent of the total cross section using the optical theorem. Other well-calculable physics processes (e.g. W/Z production, $pp \rightarrow pp + \mu^+\mu^-$) are also intended to be used for calibration.

Forward detection, close to the beam, will also be needed to expand the ATLAS physics menu with diffractive studies. A future program to study the production of central diffractive (bosonic) systems via Double Pomeron Exchange is under discussion within the collaboration, and would need precision momentum loss measurements after the first arc dipoles around $z = \pm 420$ m.

The article describes the development, construction and test of scintillating fibre prototype detectors which we tested in an electron beam at DESY in autumn 2005. After introducing the general detector concept we report about some preparatory studies of detector components and describe then the construction of the prototype detectors. This is followed by a section on the testbeam set-up and the measurement program. Section 3.2 is devoted to the data analysis and the results.

## 2. A scintillating fibre tracker for luminosity measurement

The chosen concept and design parameters are driven mainly by the following considerations. The detector must be able to detect scattered protons at the smallest possible angles (or minimum $t$ values). The detector will therefore be operated in Roman Pots very close to the LHC beam axis and should be free of any sizable ($> 100\mu$m) non-active edge region. The spatial resolution of the detector has to be significantly smaller than the spot size of the beam at the detector in order not to be limited by the detector resolution. With the proposed optics the spot size at the detector is about 130 $\mu$m, and a spatial resolution of about 30 $\mu$m, both in the vertical and horizontal plane, is therefore adequate. The detector needs to operate in vacuum (secondary vacuum inside the Roman Pot), be largely immune to the RF emitted by the LHC beam and cope with possible magnetic stray field components from a quadruole magnet in about 5 m distance. The luminosity calibrations will be performed in dedicated runs with $L \approx 10^{27}\ cm^{-2}s^{-1}$ leading to dose estimates in the 10-100 Gy/yr range. Radiation soft detector technologies can be considered as long as the detectors can be removed before LHC high luminosity operation.

### 2.1. Concept and design

A tracking detector based on scintillating fibres is able to fulfill all of the above requirements in a simple and cost effective way. Scintillating plas-tic fibers are intrinsically edgeless particle sensors. They are immune to signal pick-up from the circulating LHC beams and do not require cooling which facilitates operation under vacuum and integration in the Roman Pot. Our design (see Figs. 1 and 2) comprises scintillating plastic fibres of square cross-section ($0.5 \times 0.5$ mm$^2$) which are arranged in $UV$ or stereo geometry under an angle of $90^o$. Two layers (U and V) of 64 fibers each are glued on a 170 $\mu$m thick ceramic substrate and form a plane. Precisely machined stiffener plates of 0.5 mm thickness in the non-active regions of the plane lead to a compact and mechanically rigid assembly, a concept which has been validated by thermomechanical calculations [2]. Ten planes, staggered by multiples of $0.5$ mm$/10 \cdot \sqrt{2} = 70.7$ $\mu$m, are assembled through precisely machined hardened steel blades on precision pins to a detector with an effective fiber pitch of 50 $\mu$m. Its ultimate spatial resolution, ignoring any geometrical imperfections, is $\sigma_x = \sigma_y = 50$ $\mu$m$/\sqrt{12} = 14.4$ $\mu$m. The z-spacing of the planes is 2.5 mm. The staggering step of 70 $\mu$m per plane means that the fibre positions are aligned under an angle $\alpha = \arctan(70/2500) = 28$ mrad relative to the z-axis. To achieve optimum spatial resolution with this detector concept, the beam divergence $\sigma_{x'}, \sigma_{y'}$ must be small compared to $\alpha$. Moreover, the detector axis should be aligned with the beam axis to a precision which is again small compared to $\alpha$, e.g. to 3 mrad.

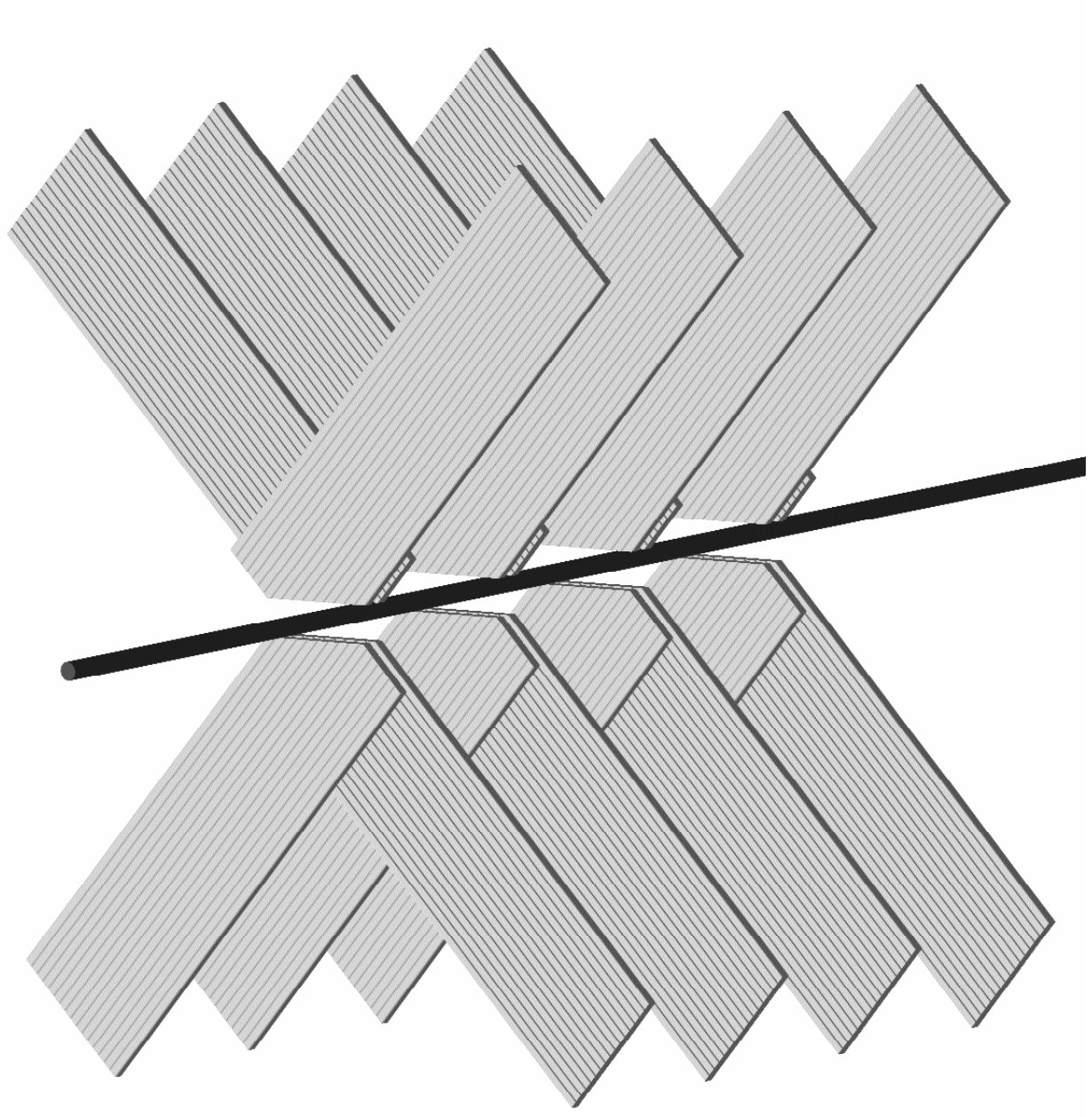

Figure 1. Principle of a scintillating fibre detector with $UV$ geometry. The active area is limited to the overlap region of the fiber layers.

[2] $-t = (p \sin\theta)^2$ is the momentum transfer squared.

Most of the fibre ends are cut at the lower end under an angle of $45^o$. The particular geometry of the Roman Pot with its thinned window requires to cut the fibres at the periphery of the layers under $90^o$. The fibres are routed over about 25 cm to a vacuum flange. Groups of 64 fibers ($8 \times 8$ fibers with a pitch of 2.3 mm) are glued into O-ring sealed connectors which fit into the flange. The scintillation light is read by 64-channels multi-anode photomultiplier tubes (MAPMT) with matched pitch [3] which are mounted on the opposite side of the flange without an optical contact medium. Custom designed high voltage divider and front-end readout electronics form a very compact assembly and sit directly on the MAPMTs.

[3]Baseline detector: Model R7600 from Hamamatsu Photonics, Japan

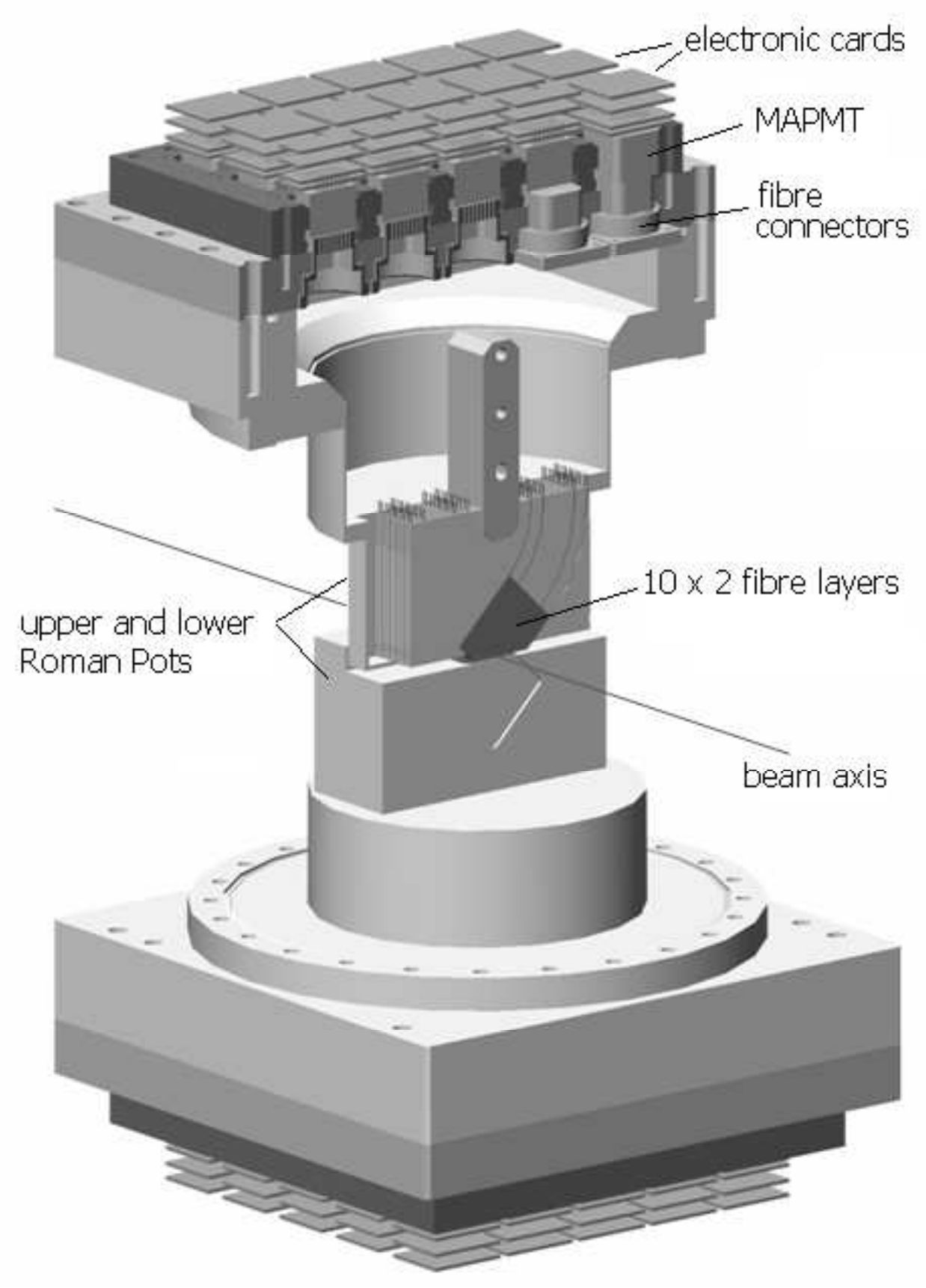

Figure 2. 3D view of an upper and lower detector and their Roman Pots. For clarity the routing of the fibres to the photodetectors is not shown.

## 2.2. Basic studies

Our baseline fibre is the single cladded 0.5 mm square fibre SCSF-78, S-type, from Kuraray[4]. Its emission spectrum expands from ~415 nm to above 550 nm with a maximum around 440 nm (for fibres of short length). Our design requires only relatively short pieces of fibres (~30 cm) but small bending radii (~30 mm). The S type fibre whose core consists of axially oriented polymer chains represents the better compromise between light attenuation length and minimum bending radius. To minimize optical cross-talk between closely packed fibres their surface is coated with a vacuum evaporated Al film of about 100 nm thickness which acts also as high quality mirror ($R \approx 0.8$). Prior to design and construction of the detector a number of lab tests were performed to characterize the fibres in terms of light attenuation length (in the emission wavelength band), minimum bending radius, scintillation light yield and surface quality of the end cut (under $45^o$ and $90^o$) [3,4]. The light attenuation length was measured by exciting the fibre with a monochromatic UV light source at 350 nm at a defined distance from the photodetector which was either a PMT with bialkali cathode or a Silicon photodiode. Light propagating in the fibre cladding was suppressed by a black coating. The results obtained with the PMT and short fibres (30 cm) are well described by an exponential with the attenuation length $\lambda_{att} = 70$ cm. Measurements with long fibres and photodiode readout can be described by two exponentials with $\lambda_{att} = 120$ cm for $0 \leq L \leq 60$ cm and $\lambda_{att} = 170$ cm for $60 \leq L \leq 300$ cm. These values remain significantly behind the >400 cm which are advertised in the Kuraray documentation for round fibres of the same type.

[4]Kuraray Co LTD., Tokyo, Japan

Special care was attributed to the surface quality of the end cuts of the fibres. As 50% of the scintillation light is emitted in the direction away from the photodetector, light reflected from the opposite fibre end can contribute significantly to the total light yield. The fibre ends were machined with a single point diamond tool on a low vibration CNC machine. High rotation frequency (20.000 $\text{min}^{-1}$) and low feed (0.02 mm per turn) gave the best results. A reflective Aluminium coating was applied by vacuum evaporation to the fibre ends which were cut under $90^o$. For fibres of 30 cm length this resulted in an increase in light yield of 58% compared to uncoated fibres. The reflectivity at the Aluminium (Al) / Polystyrene (PS) interface was found to be approximately 0.75. For fibres with a $45^o$ end cut an uncoated surface was found to lead to the highest light yield. A significant fraction of the scintillation light hits the machined surface at angles exceeding the critical angle for total internal reflection ($\theta_{crit.} = \arcsin(n_{air}/n_{PS}) \approx 39^o$). The reflected light undergoes then a reflection from the fibre side which is also Al coated. Via another reflection from the machined surface the light is reinjected into the fibre and propagates now towards the photodetector. The resulting gain in light yield was found to be about 42% (again compared to an uncoated fibre with $90^o$ end cut).

### 2.3. Construction and metrology

Six prototype detectors were built differing in geometry and fiber type. The available number of readout channels (2 MAPMTs = 128 channels) set limits for the detector sizes. A detector consisting of 10 staggered planes with $2 \times 6$ fibres each (ALFA 10_2_6) was intended for spatial resolution studies. A larger detector with 2 planes of $2 \times 32$ fibres (ALFA 2_2_32) was useful to study cross-talk between fibres and between adjacent channels on the MAPMT. Both detectors were equiped with the baseline fibres. We also built detectors with Bicron single and double cladded fibres as well as $1 \times 1$ mm$^2$ fibres from Kuraray.

The detector construction was a multi-step process which to a large part took place in a clean room (class 10.000). Whenever possible components were standardized. The detector ALFA 10_2_6 consisted e.g of 10 identical UV planes. The staggering of the planes in steps of $n \times 70$ $\mu$m ($0 \leq n \leq 9$) was only introduced when the steel blade (see below) was glued to the plane.

We used simple but precise 3-point alignment tools which were specifically designed and fabricated on a CNC machine to a precision of better than 10 $\mu$m. Optical survey[5] with an accuracy of about 3 $\mu$m was performed at three stages of the fabrication process. The sequence of steps was:

1. LASER cutting in industry of the central and stiffener plates from $Al_2O_3$ ceramic sheets at a modest precision of about $\pm 30$ $\mu$m.
2. Survey of the plates' geometry and selection

[5] CERN metrology service, 3D optical coordinate measurement machine MAHR OMS-600

of the best matching sets.

3. Gluing of left stiffener plates on front and rear side. Alignment via lateral and bottom edges.
4. Survey for control and determination of reference points.
5. Positioning and glueing of the Al coated fibres, one side after the other. The fibres were aligned with the $45^o$ edge of the left stiffener plate and gently compressed by the right plate. 'Optical cement' Bicron BC-600 was used to glue both fibres and ceramic components. The ends of the fibres with a $90^o$ cut had been machined and Al-coated before. The fibres with the uncoated $45^o$ cut were left a few mm longer than needed.
6. Machining (single point diamond tool) of the fibre ends ($45^o$ cut) to final length and surface quality.
7. Glueing of the support blade made of hardened steel. The position of the blade was chosen to produce the desired staggering. This step was done by means of an alignment plate which provided a specific position for each displacement and allowed also to (partly) correct for small geometrical errors accumulated in the sequence up to this point.
8. Survey of fibre positions and angles relative to the precision holes of the steel blade.
9. Assembly of the planes on a support arm through two precision pins.
10. Threading of the fibres into the connector and subsequent glueing. A system has been adopted which avoids that adjacent fibres on the detector are not connected to adjacent channels on the MAPMT.
11. Machining (diamond tool) of the protruding fibre ends on the connector.

All prototype detectors were tested in a lab set-up where the fibres were exposed to a Sr-90 source and read out by a MAPMT. These tests [5] gave promising indications of light yield and uniformity of the detectors.

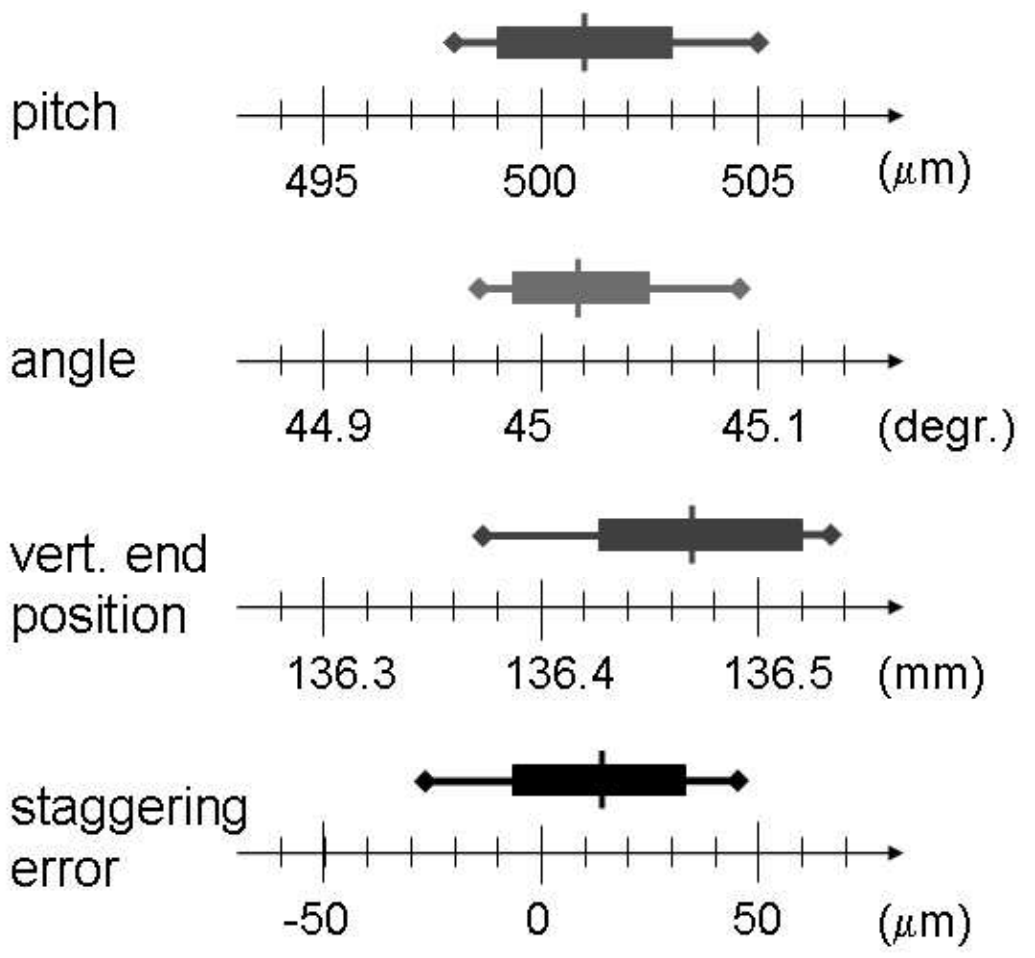

Figure 3. Results of metrological analysis of the 20 layers of the ALFA 10_2_6 detector. The bars show the ± RMS error, the lines mark the minimum and maximum values of the 20 measured values. The absolute values of the vertical end positions of the fibres are not relevant, only their spread is of importance.

The results of the final survey of the detector ALFA 10_2_6 which was used for the spatial resolution measurements are shown in Fig. 3. The geometrical data of all detectors like the positions of the first fibre (x,y,z) of each layer, the average fibre pitches and angles of each layer were recorded in *geometry files* and later used in the space point reconstruction (see section 3.2.4).

## 3. Test of prototype detectors in a 6 GeV electron beam at DESY

The main goals of the test beam in the DESY T22 area were to prove the validity of the detector concept and to study in particular the following characteristics: light yield ( more precisely the photoelectric yield) of the fibres, cross-talk effects on various levels, detection efficiency, track reconstruction efficiency, spatial resolution and edge sensitivity.

### 3.1. Experimental set-up

The DESY T22 zone provides a low intensity electron (or positron) beam ($\leq$ 1 kHz) of up to 7 GeV energy. The electrons or positrons are converted bremsstrahlung photons from carbon fiber targets in the lepton synchrotron DESY II. The beam is collimated to a spot of about 1 $cm^2$ and has low divergence $\sigma_x \approx \sigma_y \approx 1$ mrad. The detectors were mounted on a motorized XY table which allowed to centre them w.r.t. the beam. The beam trigger was defined by a coincidence $S_1 \times S_2 \times S_3$ of three 'finger' scintillators of $9 \times 9$ $mm^2$ size. A beam telescope [6] consisting of three silicon microstrip detector stations (1 station before, 2 stations behind the detector under test) was used to predict the trajectory of the incident electrons. The active area of the telescope was $32 \times 32$ $mm^2$. The strip readout pitch of 50 $\mu$m resulted in a spatial resolution of $\sigma_x \approx \sigma_y \approx 20 - 30$ $\mu$m, depending on the amount of additional material in the beam.

The MAPMTs of the ALFA detectors were read out in two different ways. High gain operation ($G \approx 5 \cdot 10^6$ at HV = 950 V) allowed to feed the signals via a twisted pair cable directly into VME based charge sensitive ADCs[6]. Alternatively the signals were amplified, shaped and eventually discriminated by the OPERA front-end chip [7,8]. The MAPMTs were then operated at typically half the above gain value. Custom designed readout cards housing two OPERA chips (32 channels/chip) were plugged directly on the output pins of the MAPMTs. The discriminated signals were handled by a FPGA mounted also on the cards and sent via USB link to a PC. The OPERA chip provides the possibility to access also the charge amplitude of each channel. The analogue signals were converted by the FADC of a digital oscilloscope[7] and read by the same PC over a GPIB interface. The analogue readout electronics (3840 channels) of the Silicon telescope was based on Viking VA frontend chips[8], read out in a serial way over VME based sequencer and ADC[9]. Zero suppression was applied to the telescope data. All VME data (Si-telescope and ALFA) was transferred over an optical link[10] to a data acquisition PC. A classical trigger and busy logic was implemented with standard NIM modules. The data taking rate was limited by the conversion time of the ADCs of the Si-telescope to about 70 Hz.

The raw data from ALFA and Si-telescope was jointly stored in PAW Ntuples. A set of specific analysis routines was implemented in the TESA [9] package, a set of Fortran and PAW routines developed previously by the ZEUS microvertex detector team.

### 3.2. Analysis and Results

#### 3.2.1. Photoelectric yield

A sufficiently high light or photoelectric yield is a key requirement for good detection efficiency and finally spatial resolution. The expected photoelectric yield

$$N_{p.e.} = N_{scint} \cdot \epsilon_{acc} \cdot \epsilon_{transp} \cdot \epsilon_{refl} \cdot \epsilon_{gap} \cdot \epsilon_{Q_{eff}} \quad (1)$$

can be estimated with $N_{scint} = 1660$ $mm^{-1} \cdot 0.48$ mm $= 797$ being the number of generated scintillation photons for a traversing m.i.p., $\epsilon_{acc} = 0.042$ being the geometrical acceptance factor of a rectangular fibre, $\epsilon_{transp} = exp(-30/70) = 0.65$ being the transport efficiency due to optical absorption, $\epsilon_{refl}$ =1.58 ($90^o$ cut) or 1.42 ($45^o$ cut) being the gain due to reflection from the opposite fibre end, $\epsilon_{gap} = 0.9$ being the transmission at the fibre / air / glass interface without any grease, and $\epsilon_{Q_{eff}} = 0.14$ being the effective quantum efficiency of the MAPMT[11]. The equation leads to $N_{p.e.} = 4.3$ for fibers with a $90^o$ cut and 3.9 for fibers with a $45^o$ end cut.

A photoelectric yield of 4 promises an excellent single fibre detection efficiency. An optimistic estimate can be derived from $\epsilon_{det} \approx 1 - P(0,4)$ where $P(0,\mu) = e^{-\mu}$ corresponds to the Poissonian probability to have zero photoelectrons when the average number is $\mu$. From $\mu = 4$ follows a single fiber efficiency $\epsilon_{det} = 98.2\%$. This simple estimate ignores geometrical inefficiencies (cladding, glue between fibres) and assumes that a single photoelectron can be detected by the DAQ system with 100% efficiency.

[6]Model V792 from CAEN, Viareggio, Italy
[7]LeCroy WaveRunner LT344
[8]IDEAS, Snaroya, Norway
[9]CAEN V551B (C-RAMS) and V550 (sampling ADC)
[10]PCI to VME link SIS1100/3100, Struck SIS GmbH, Hamburg, Germany
[11]$\epsilon_{Q_{eff}}$ is the product of quantum efficiency at 450 nm ($\approx 0.2$) and photoelectron collection efficiency. The latter has not been measured explicitly. We took the value of 0.7 which was communicated by M. Metzger, Hamamatsu Photonics, Switzerland, for the R7600-00-M64 MAPMT.

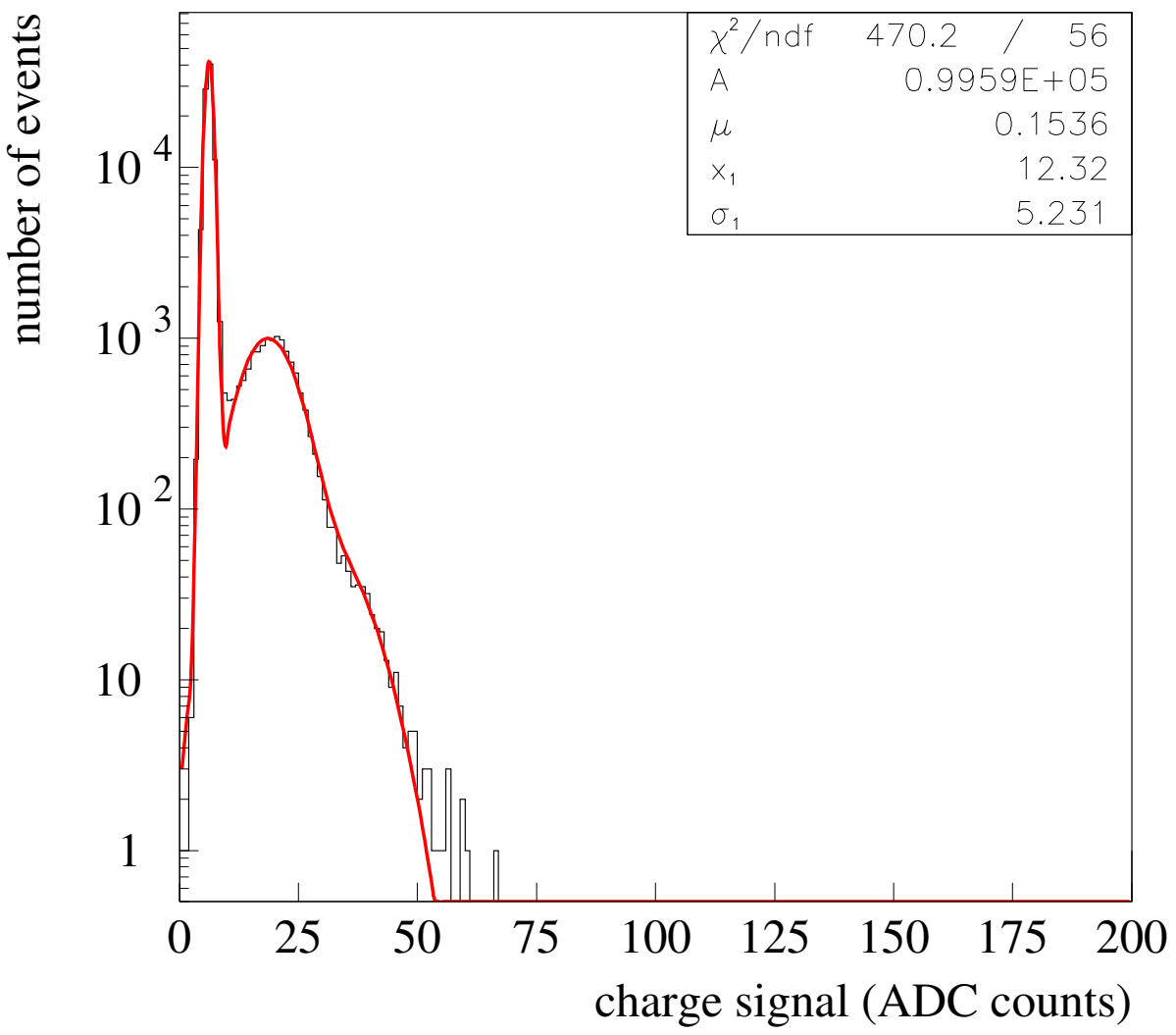

Figure 4. Single photoelectron spectrum obtained at 950 V. The onset of the second photoelecton is visible as shoulder around 33 ADC counts.

To calibrate the charge amplitude of the MAPMTs, all channels of both photodetectors were directly exposed to very low light levels from a pulsed blue LED. The charge spectra recorded with the ADC modules for different HV values (900, 950, 1000 V) were then fitted (`MINUIT`[10]) with a Poisson distribution convoluted with a Gauss function

$$Q(x) = A \sum_{0\leq i\leq n} P(i,\mu)\cdot e^{-(x-x_i)^2/2\sigma_i^2} \qquad (2)$$

The variable $x$ is the charge amplitude in ADC counts. The fit provides apart from the overall normalization $A$ the average number of photoelectrons $\mu$, the position of the pedestal $x_0$ and photoelectron distributions $x_i$, with $x_i = i\cdot x_1$, and their width $\sigma_i$, where $\sigma_i^2 = \sigma_0^2 + i\cdot\sigma_1^2$, with $\sigma_0$ being the width of the pedestal peak. Fig. 4 shows a typical single phototelectron spectrum and the corresponding fit. The single photoelectron and pedestal peak are well separated.

The charge spectra obtained with the scintillating fibres in the electron beam, see Fig. 5, could then be fitted using the above equation while the calibration parameters $x_1$ and $\sigma_1$ were kept on the values determined by the calibration. In principle the overall normalization $A$ and the photoelectric yield $\mu$ are the only fit parameters. A further fit parameter $A_0$ was added to allow for an independent normalization of the pedestal peak, because not every fibre was hit in each recorded beam event.

The fit led to satisfactory results, however it failed to describe a small obvioulsy non-Poissonian excess of primarily single and very few double photoelectrons which was consistently present in the spectra. This excess was attributed to cross-talk from adjacent MAPMT channels (discussed in detail below) and could be incorporated in the fit function by simply adding a second Poisson distribution with a small $\mu_{CT}$ parameter

$$Q'(x) = Q(x,\mu) + Q_{CT}(x,\mu_{CT}). \qquad (3)$$

This fit function describes the data well and achieves values of $\chi^2/n.d.f.$ of about 2.3. The detected number of photoelectrons for our baseline fibre was 3.93 ± 0.18 (averaged over 40 fibres with a $45^o$ cut) and 4.45 ± 0.50 (averaged over 24 fibres with a $90^o$ cut). The agreement with the expected values is better than 5%. The values are reasonably uniform over a fibre plane, min/max = 3.4/4.2 for $45^o$ and 3.6/5.0 for $90^o$, respectively, and do not depend on the applied HV of the MAPMT. The relative cross-talk contribution defined as $\mu_{CT}/(\mu+\mu_{CT})$ is 3.4±1.3%. It describes the signal fraction which a given MAPMT channel received from the 8 adjacent channels. It depends therefore not only on the strength of the cross-talk (discussed below) but also on the probability that an adjacent channel was hit. Within the errors the cross-talk contribution was found to be independent of the HV.

In addition to the detectors with baseline fibres (Kuraray SCSF-78, 0.5 × 0.5 mm$^2$), two detectors were equiped with fibres of larger section (1 × 1 mm$^2$) and with Bicron square fibres of 0.5 × 0.5 mm$^2$ size, single (BC 612) and double cladded (BC 612 MC) types. Data was also taken with different external mirrors (rigid and soft) mounted under the fibre ends cut at $45^o$. A further study dealt with the thickness of the air gap between the fibre connectors and the MAPMT entrance window. The results obtained for the different fibre types and geometries are summarized in table 1.

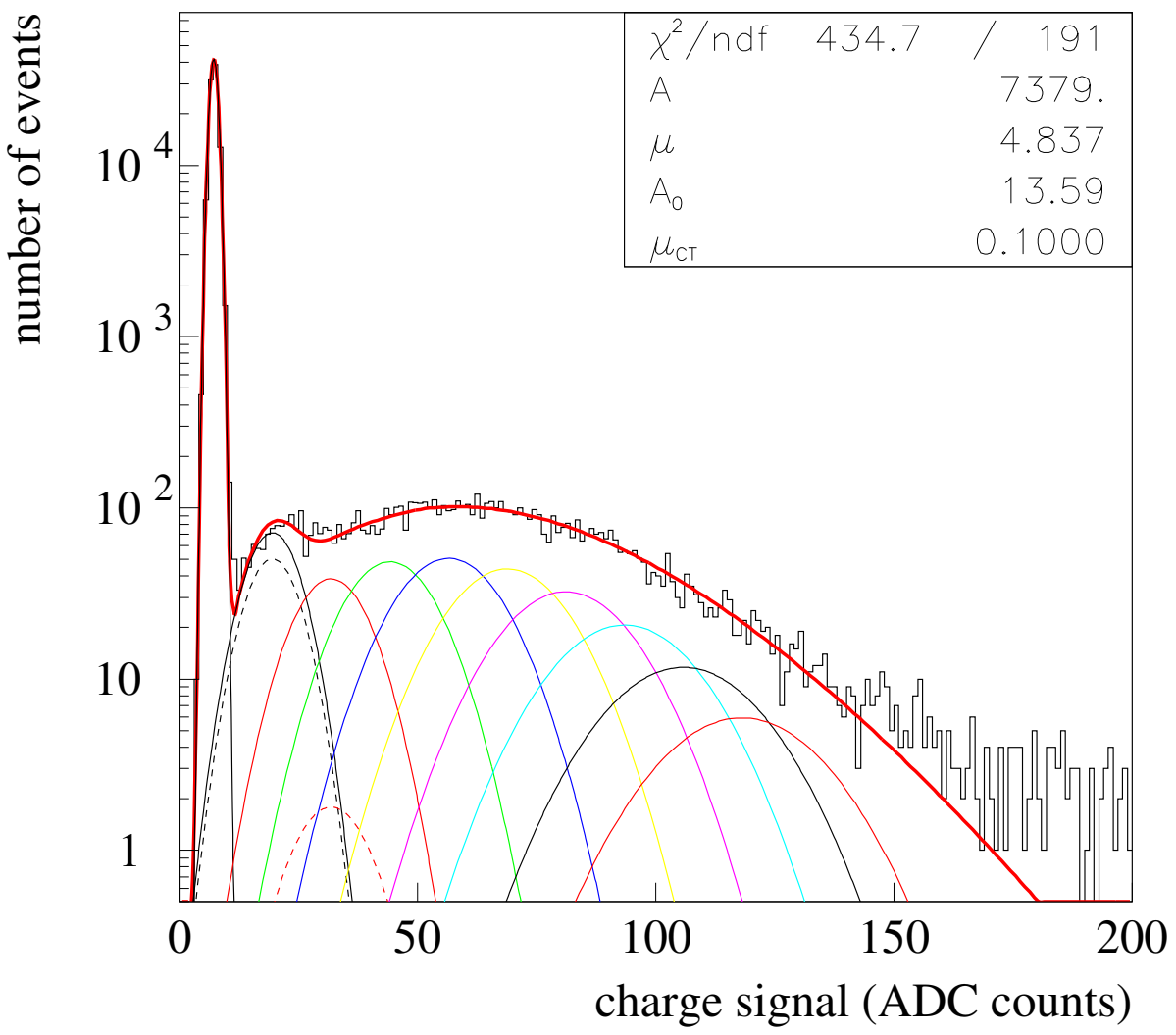

Figure 5. Photoelectron spectra from a 0.5 mm fibre ($45^o$ cut) obtained at 950 V. The average number of photoelectrons is 4.8, the cross talk term contributes additional 0.1 photoelectrons. The Gauss curves represent the $i = 1...9$ photoelectrons contributions. The dashed Gauss curves show the cross-talk term.

Table 1
Results of the fits of the signal spectra at HV = 950 V: Photoelectric yield and cross-talk. The errors are RMS values.

| fibre type | $N_{p.e.} (= \mu)$ | cross talk contribution (%) |
|---|---|---|
| baseline $45^o$ | $3.93 \pm 0.18$ | $3.4 \pm 1.3$ |
| baseline $90^o$ | $4.45 \pm 0.50$ | $3.4 \pm 0.8$ |
| baseline mirror $45^o$ | $4.87 \pm 0.49$ | $4.2 \pm 1.8$ |
| baseline $45^o$ 0.5 mm air gap | $4.09 \pm 0.25$ | $6.89 \pm 2.25$ |
| baseline $45^o$ 1 mm air gap | $3.88 \pm 0.25$ | $8.6 \pm 2.4$ |
| BC 612 $45^o$ | $3.02 \pm 0.36$ | $3.3 \pm 2.1$ |
| BC 612 MC $45^o$ | $3.20 \pm 0.32$ | $3.6 \pm 1.8$ |
| SCSF-78 $1 \times 1$ mm$^2$ $45^o$ | $7.39 \pm 0.44$ | $4.3 \pm 2.6$ |

### 3.2.2. Cross-talk

We observed two phenomena which gave rise to cross-talk effects between channels. The first is caused by optical and 'electron-optical' cross-talk. Propagation of scintillation light from one fibre to another on the level of the detector was efficiently supressed by the opaque Al coating. However, optical cross-talk with the adjacent MAPMT channels can arise from multiple reflections in the entrance window of the MAPMT and by direct emission of a small fraction of the light into an adjacent MAPMT channel. The latter is due to light diffusion at the fibre end cut in combination with the expansion of the light cone emitted from the fibre in the 0.8 mm thick MAPMT window. A small amount of electron-optical cross-talk ($\ll$ 1%) is due to charge sharing during the avalanche formation in the MAPMT dynode structure. These effects lead to cross-talk signals which are small compared to the actual signal. In our case, where the proper signals were around 4 photoelectrons, the cross-talk signals produced in this way consisted essentially of single photoelectrons. The optical cross-talk probability was determined by selecting hits in an arbitrary MAPMT channel which passed a $5\sigma$ pedestal cut. Considering those as central hits the average numbers of photoelectrons $\mu$ in the adjacent 8 MAPMT channels (direct and diagonal) were determined. Dividing these values by the one of the central fibre led to the cross-talk probabilities. For the baseline fibre they were found to be 1.3% for the direct and 0.4% for the diagonal neighbours.

The second phenomenon manifested itself as symmetric cross-talk from the hit fibre to the two or three neighbours in the same fibre layer. The signal spectra in the adjacent fibres showed however amplitudes which were comparable to the amplitude in the central fibre. The abundance

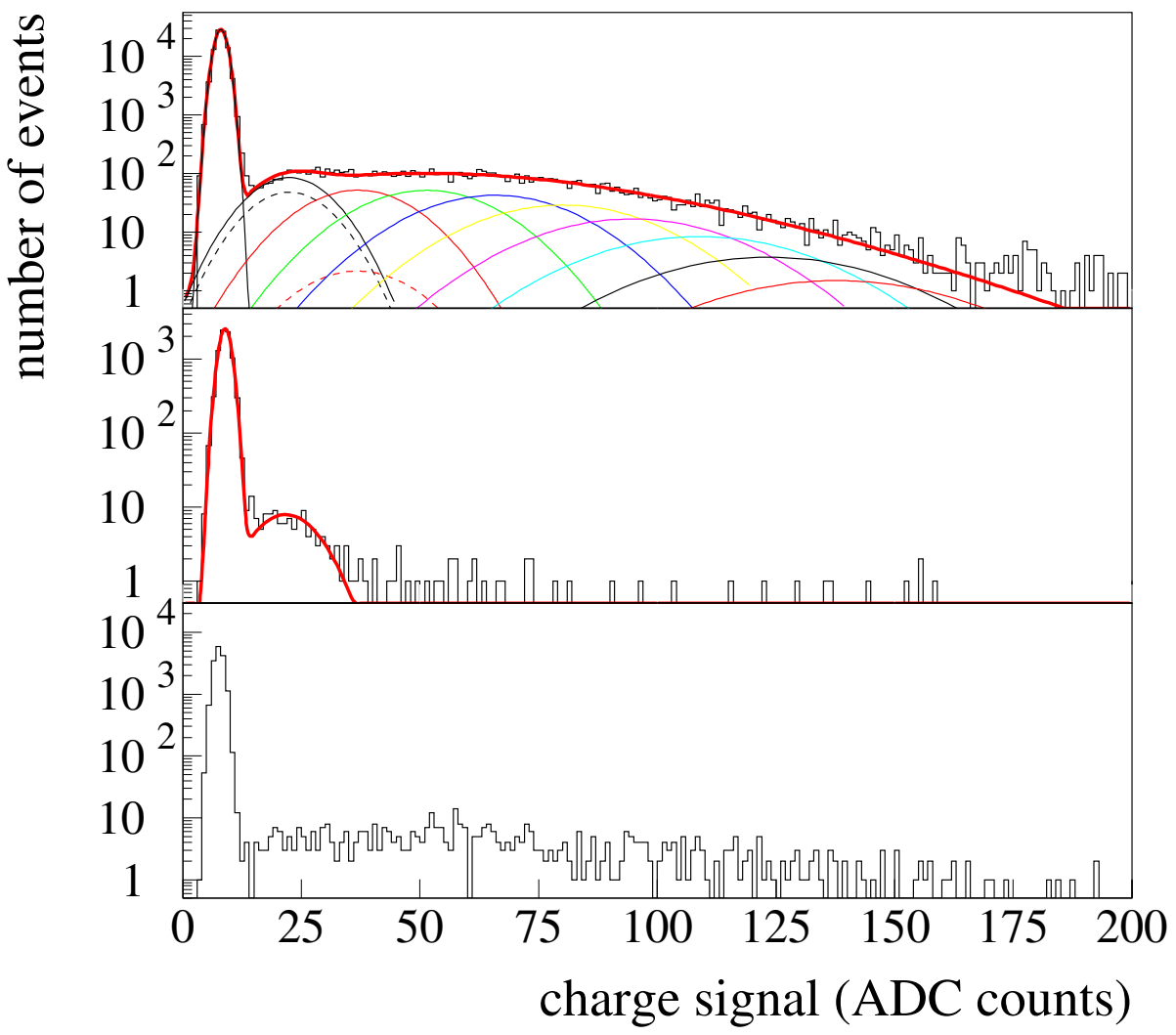

Figure 6. Comparison of typical signal spectra. Normal hits (top), optical (middle) and physics cross-talk (bottom).

Table 2
Results of the cross-talk (CT) study. Cross-talk probabilities in %. The uncertainty of these results is typically ±0.5% (absolute).

| fibre type | optical CT[1)] | physics CT[2)] |
|---|---|---|
| baseline | 1.3, 0.4 | 3.7, 1.3, 0.9 |
| baseline 0.5 mm air gap 45$^o$ | 1.4, 0.5 | 4.1, 1.4, 1.0 |
| baseline 1 mm air gap 45$^o$ | 2.4, 0.8 | 4.4, 1.3, 1.2 |
| SCSF-78 1.0 mm | 1.6, 0.6 | 4.4, 2.2, -[3)] |
| BC 612 0.5 mm | 1.0, 0.4 | 2.8, 0.7, 0.5 |
| BC 612 MC 0.5 mm | 1.0, 0.4 | 1.8, 0.4, 0.3 |

[1)] The two values correspond to the direct and diagonal neighbours on the MAPMT. [2)] The values are for the first, second and third neighbouring fibres in the same layer. [3)] Layer has only 6 fibres in total.

of these 'satellite' hits decreased rapidly with the distance from the central fibre. We attributed this *physics* cross-talk to hits by $\delta$-electrons which were produced by the incident electron upstream in the set-up. We determined these physics cross-talk probabilities by again selecting first hits in a central fibre (5$\sigma$-cut), measured then the number of hits in the 3 adjacent fibres on both sides which passed the same cut and calculated finally the hit ratios. For the baseline fibres the three next neighbours on both sides showed the following average probabilities: 3.7%, 1.3% and 0.9%.

Typical signal spectra illustrating normal hits, optical and physics cross-talk are compared in Fig. 6. The results of the analysis for the different fibre types and geometries are compiled in Table 2.

### 3.2.3. Detection Efficiency

The efficiency of individual fibres was directly determined by selecting with the Si-telescope events which have hits in the central area of the fibre. The studies described in this section were performed with the ALFA 10_2_6 detector, equiped with baseline fibres. Events within a margin of 50 $\mu$m from the fibre edges were rejected to ensure containment in the sensitive core of the fibre. The efficiency was studied as function of the hit-defining cut, applied to the signal amplitude calibrated in photoelectrons. Typical examples of the single fibre efficiencies are shown in figure 7. The observed efficiency was about 98% for a cut of zero p.e., as anticipated from the photoelectric yield, and dropped to about 92% for a cut of 0.9 p.e., which has been applied for the track reconstruction studies described in 3.2.4. In order to investigate potential efficiency losses near the fibre edges, which could originate from insensitive cladding and glue remnants between fibres, a horizontal position scan was performed accumulating data in slices of 50 $\mu$m width. The efficiency was then evaluated in each of the slices, both for an entire plane consisting of 6 fibres and for individual fibres. The result for a fixed cut at 0.9 p.e. is shown in figure 8. The plane efficiency was rather uniform and reached 95% in the core of the fibres. A small drop to 80% efficiency in its deepest point occured in the inter-fibre gap, which spans about 150 $\mu$m between the plateau region of adjacent fibres.

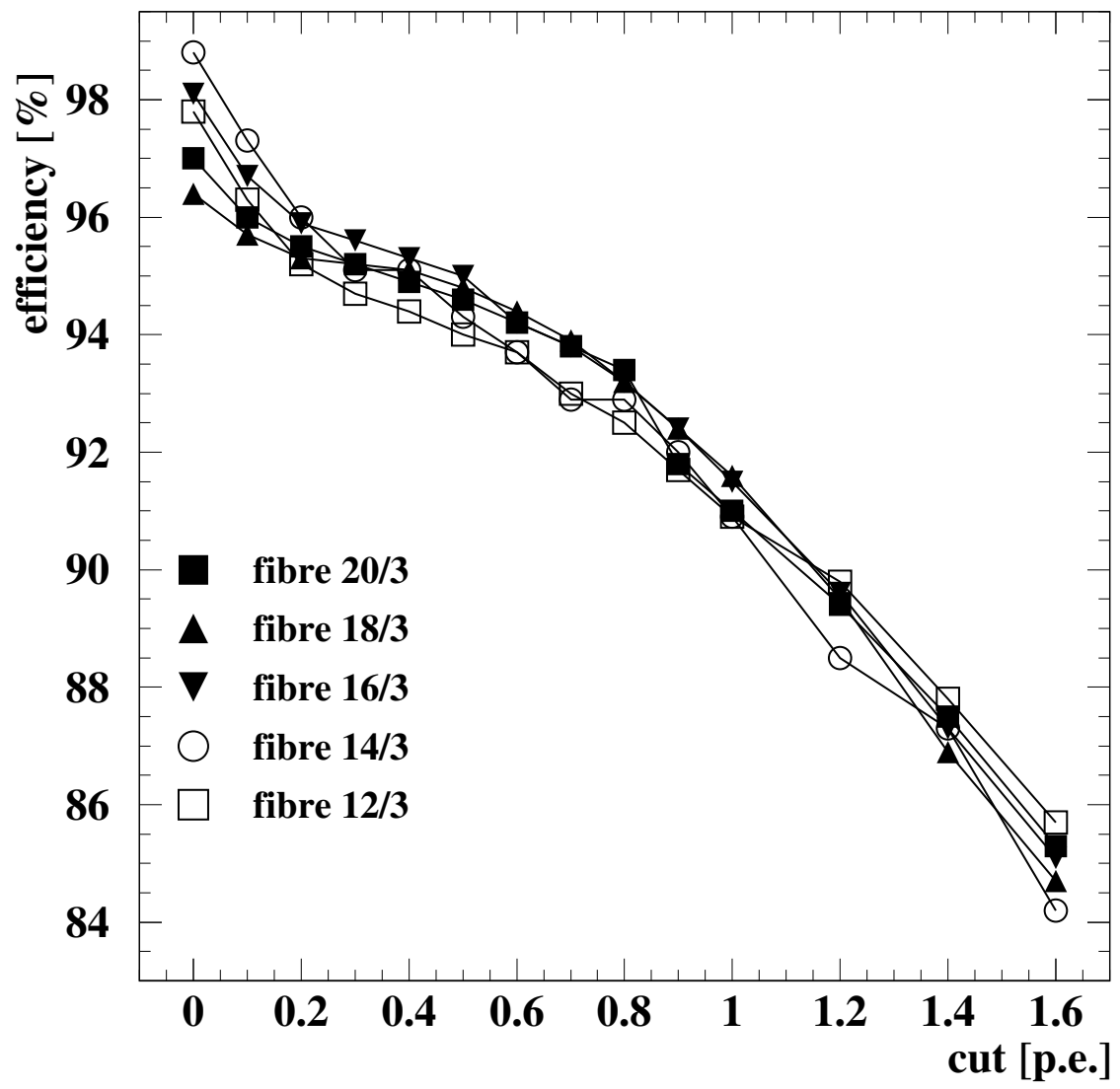

Figure 7. Single fibre efficiency as function of the cut in units of photoelectrons for typical fibres.

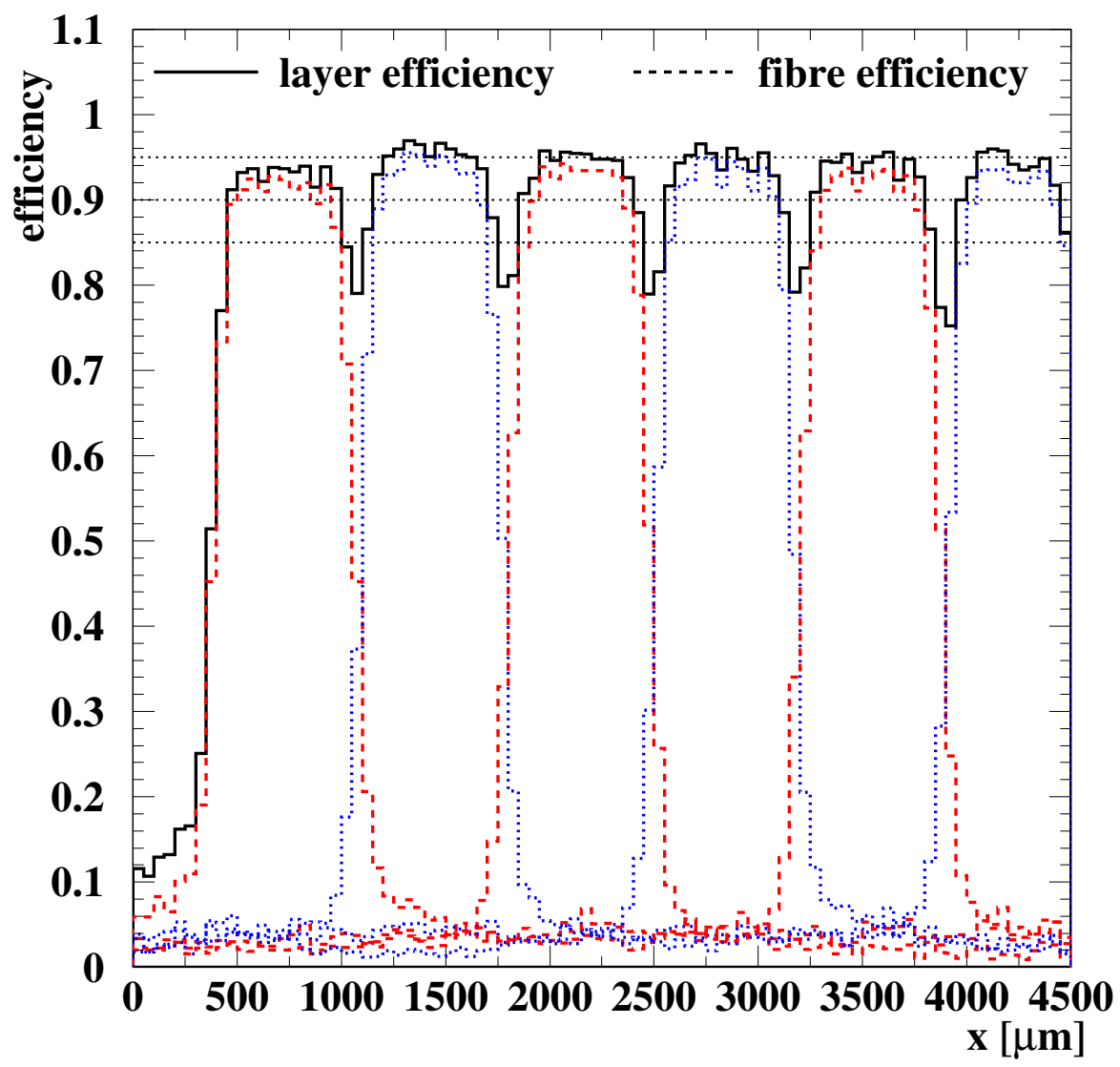

Figure 8. Layer efficiency (solid line) and fibre efficiency (dashed line) for a fixed cut of 0.9 p.e as function of the horizontal beam position.

The total detection efficiency, i.e. the probability to have at least one hit in any plane, was 99.9%.

### 3.2.4. Space point reconstruction

The main purpose of the ALFA detector is the determination of the $t$ spectrum of elastically scattered protons through the measurement of their impact points $(x, y)$ in the transversal plane. The tracking studies described in the following were carried out mainly for ALFA 10_2_6. The space points were determined from the crossing area of hit U- and V-fibres. The resolution was achieved by combining hits from the ten longitudinal planes, each of them staggered by multiples of 50 $\mu$m$\times\sqrt{2}$.

The impact point reconstruction proceeded in several steps. First hits were defined by requesting the calibrated signal amplitude to be above a certain cut, which is set to 0.9 p.e. unless stated otherwise. Ideally a charged particle traversing ALFA perpendicularly would hit 20 fibres. The observed hit multiplicity distribution, shown in figure 9, has on average a slightly higher number of hits. On one side more than one fibre can be hit per layer, resulting for example from $\delta$-electrons, on the other side inefficiencies can lead to layers without any hit, as illustrated in figure 10. On average 18 layers had at least one hit and only 14% of the events had all 20 layers hit, while a significant tail extended toward lower multiplicities. The hit multiplicity per layer is shown in figure 11. On average a layer had about 80% of single hits, 10% of zero hits and 10% of multiple hits. The presence of several hits in a plane complicated the track finding, as the right hit generated from the primary particle had to be selected and hits originating from secondary interactions or cross talk had to be rejected.

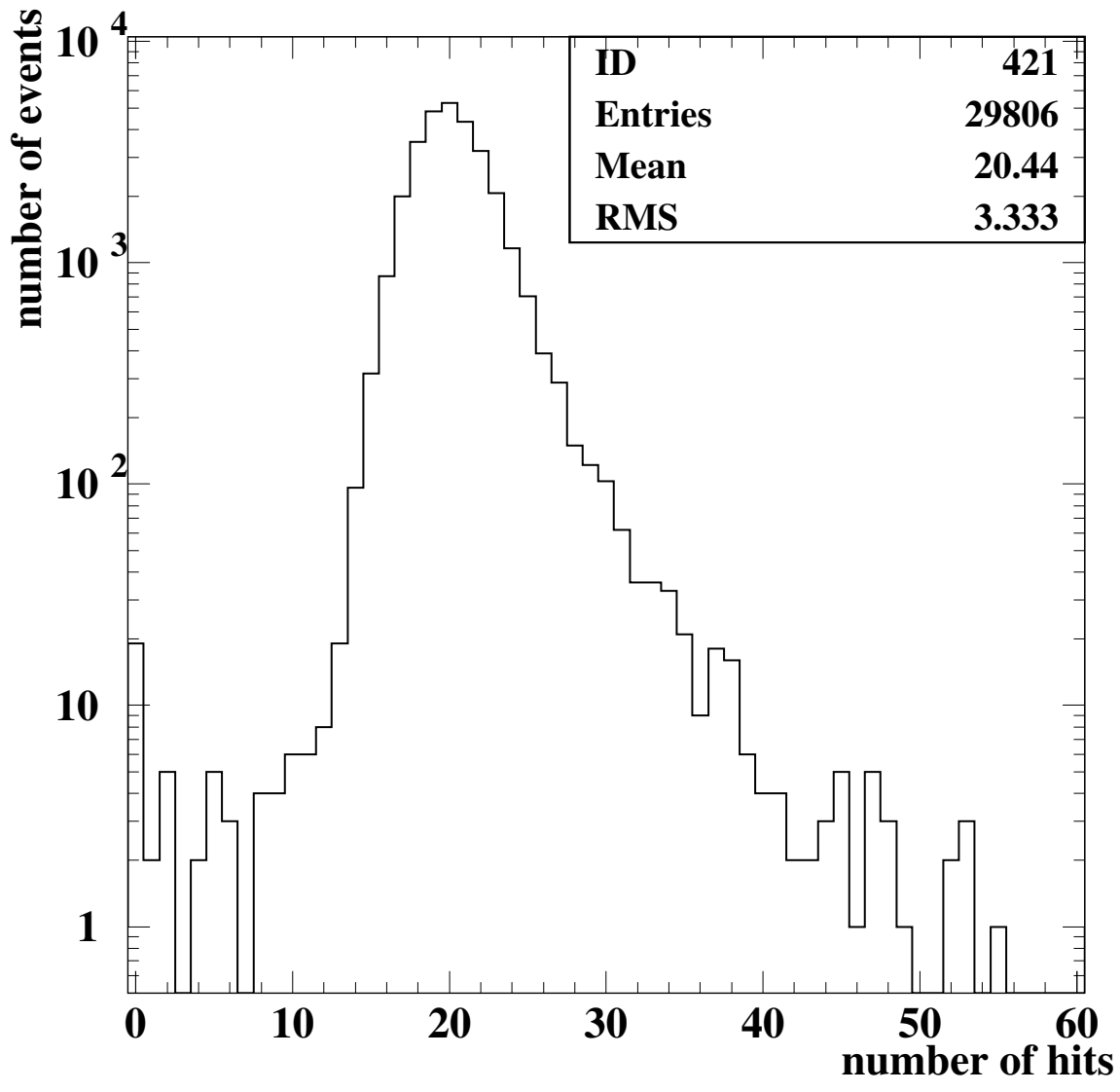

Figure 9. Hit multiplicity distribution in ALFA 10_2_6 for 6 GeV electrons.

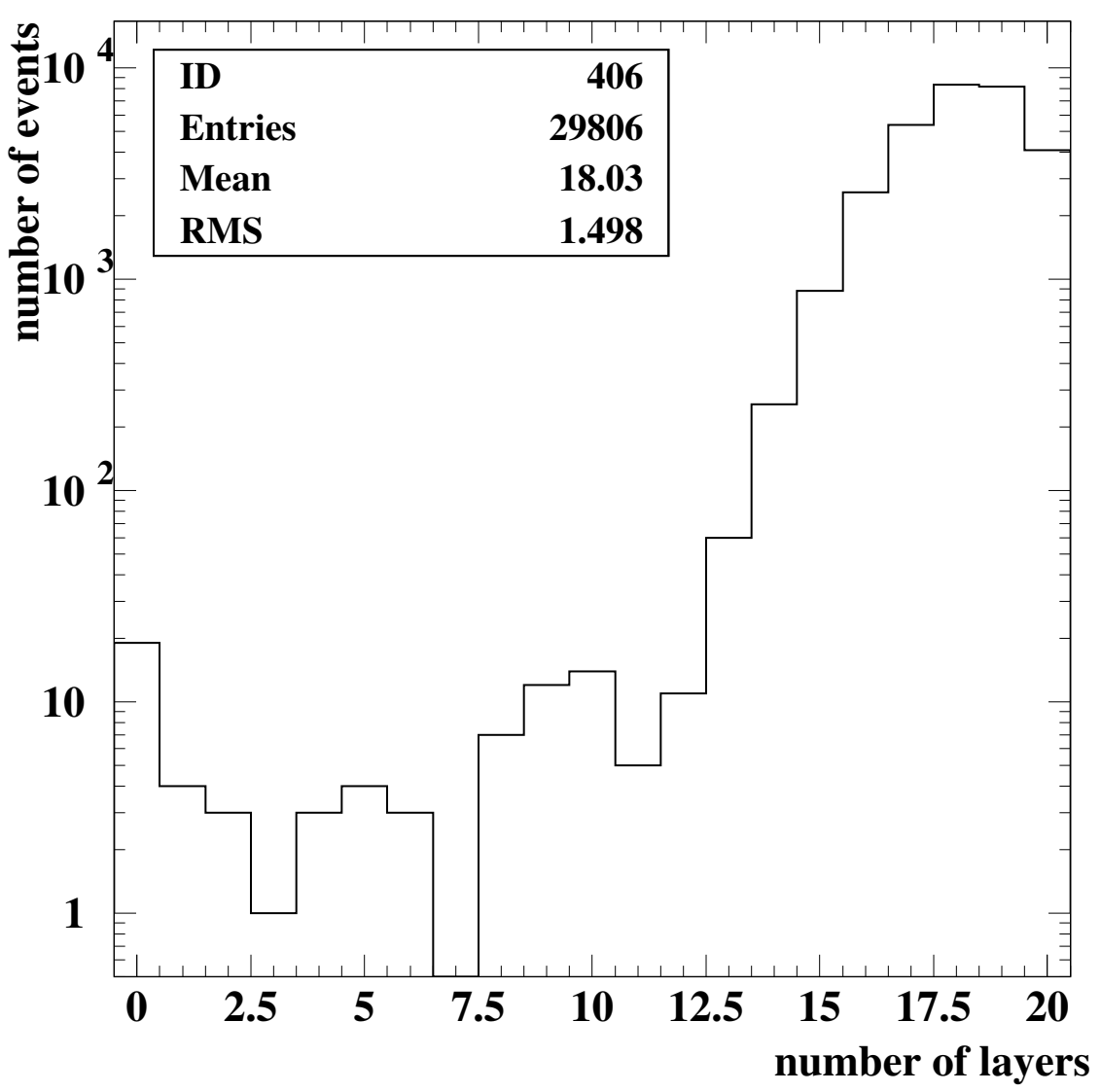

Figure 10. Layer multiplicity distribution in ALFA 10_2_6.

In the second step of the space point reconstruction a track seed was determined using only planes with exactly one hit. The seed track was determined with the *minimal overlap algorithm*, which calculates for U- and V-fibres separately the geometrical area common to the selected fibres, i.e. the overlap area resulting from the staggering of the planes. This procedure used again a seed as starting point, which was taken from the centre of gravity of the hits. In this way fibres far away from the seed which had no overlap with the other fibres were discarded. The centre of the overlap area represented by a linear equation was finally used to calculate the crossing point of U- and V-fibres, which yielded the space point seed $(x, y)$. In order to quantify the quality of a track a metrics $\Delta$ was introduced, which measured the distance of closest approach of the centres of the fibres to the space point candidate, averaged over all participating fibres.

In the third step the minimal overlap algorithm was executed in a loop over layers with more than one hit, iterating over all hits in that layer. The hit yielding the best value of $\Delta$ was retained, provided the current $\Delta$ was smaller than the corresponding seed value. At the end of the iteration the final space point was hence based on the hit configuration yielding the highest quality track.

The space point resolution of ALFA was determined from a comparison of the reconstructed $x$ and $y$ values with the tracks measured by the telescope. The coordinate systems of the Si-telescope and of ALFA were aligned ($x, y$ and rotation around the beam axis) allowing for individual alignment constants for each layer of fibres, starting from the parameters provided by the geometrical survey. Prior to this alignment procedure an internal alignment procedure of the three telescope stations was performed, which also yielded an estimate of the intrinsic telescope resolution by comparing the track impact predicted by two stations with the measured impact in the third.

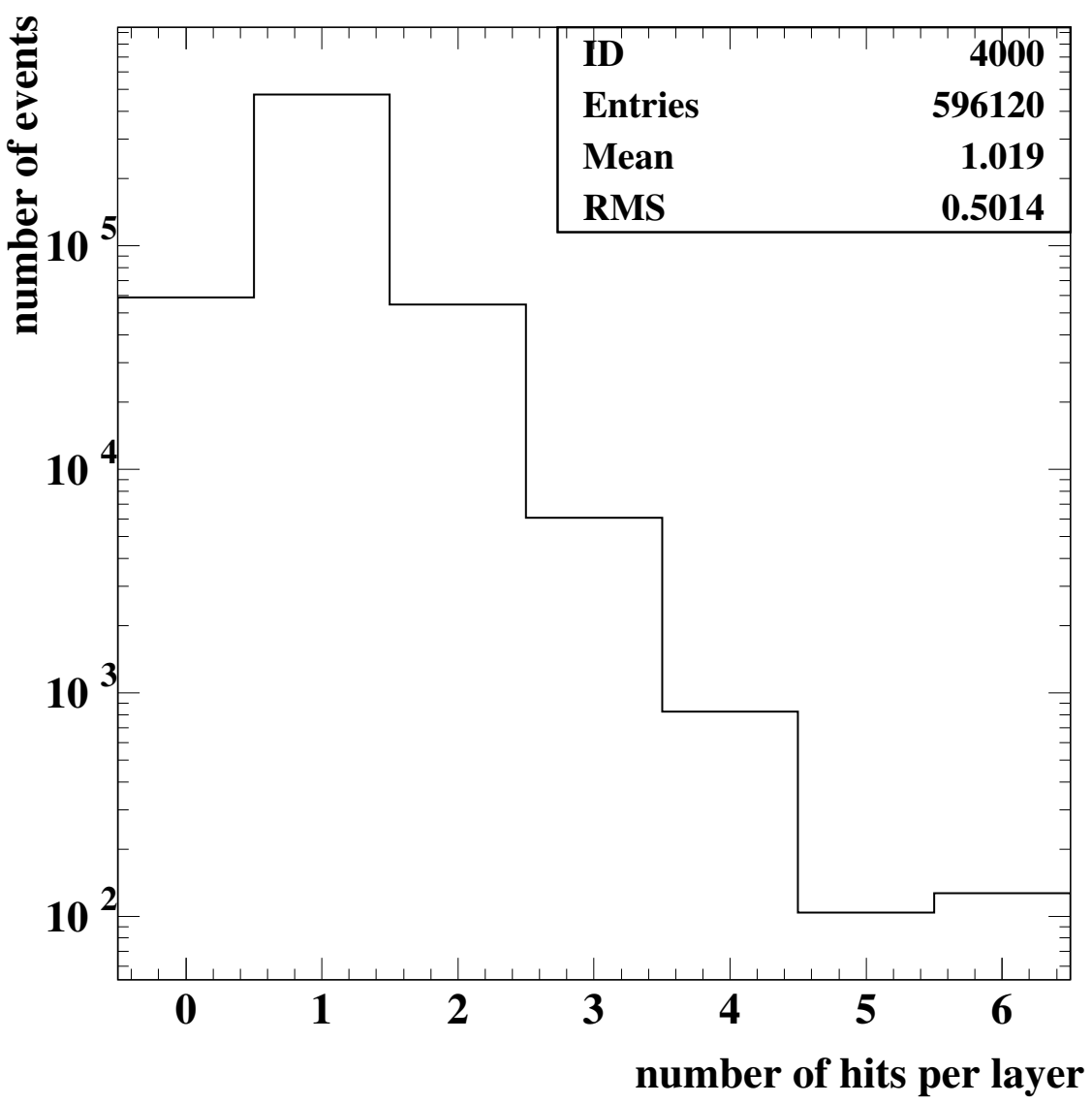

Figure 11. Hit multiplicity distribution per layer, averaged over all layers.

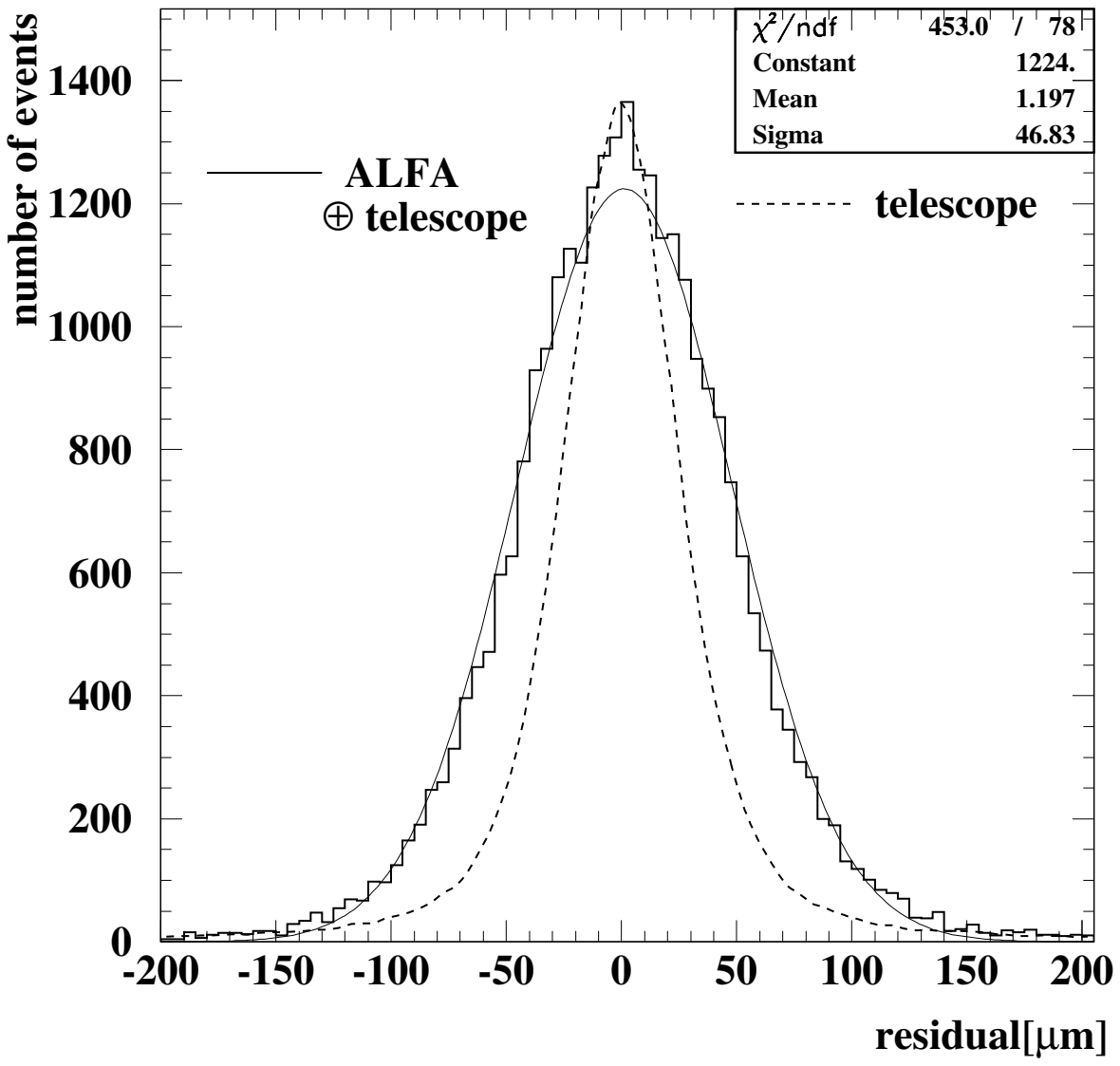

Figure 12. Detector ALFA 10_2_6. The residuals $x_{\mathsf{ALFA}} - x_{\mathrm{telescope}}$ (solid line) compared to the intrinsic resolution of the telescope (dashed line).

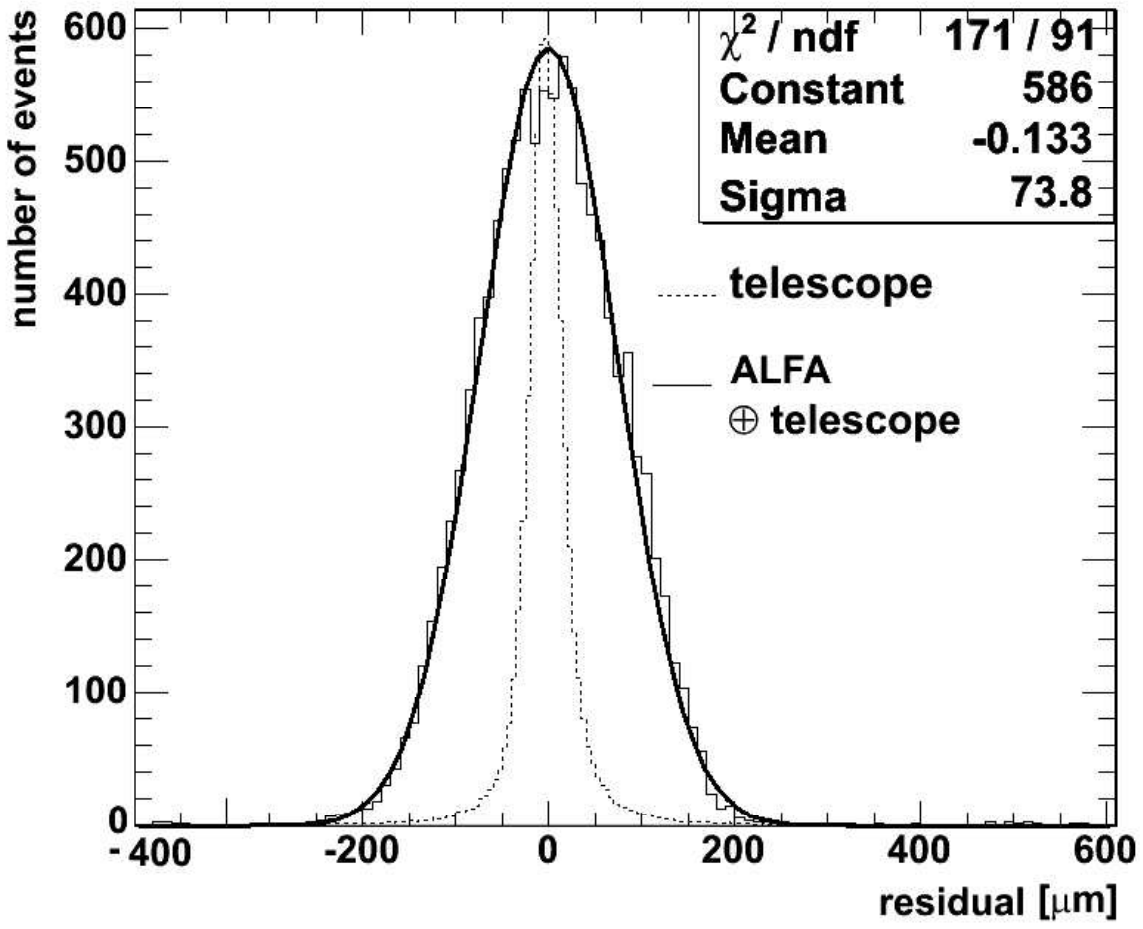

Figure 13. As previous figure, however for the detector ALFA 2_2_32.

For the determination of the ALFA resolution events were selected with the telescope in the centre of the detector within a circle of 1 mm radius and with polar angle below 1 mrad. The difference of the $x$-position reconstructed with ALFA 10_2_6 and with the telescope is shown in figure 12, the plot is very similar for the $y$-coordinate. The corresponding data obtained with the detector ALFA 2_2_32 is shown for comparison in Fig. 13. ALFA 2_2_32 consists of only two layers with 2 × 32 fibres each, resulting in an effective pitch of 250 $\mu$m and a theoretically achievable resolution of 250 $\mu\mathrm{m}/\sqrt{12} = 72$ $\mu$m.

The distribution of the residuals $x_{\mathsf{ALFA}} - x_{\mathrm{telescope}}$ was fitted by a Gaussian in order to determine the total resolution. In the same way the intrinsic telescope resolution was determined, using the residuals from the central station, as shown in Figure 12. The expected resolution for ALFA alone was obtained from the quadratic subtraction of the telescope contribution from the total resolution. The results on the resolution are summarised in table 3. The space point reconstruction efficiency was 99 %, the 1% inefficiency was caused by events with insufficient single-hit planes to get a decent track seed.

Table 3
Space point resolution of ALFA 10_2_6 for 6 GeV electrons.

| coordinate | resolution [$\mu$m] |
|---|---|
| $x$ total | 46.83 |
| $y$ total | 46.98 |
| $x$ telescope | 29.95 |
| $y$ telescope | 29.50 |
| $x$ ALFA | 36.00 |
| $y$ ALFA | 36.56 |

The ALFA tracking performance showed little sensitivity to variations of the hit-defining cut, as illustrated in figure 14. Also the track reconstruction efficiency was little affected over the considered cut variation range.

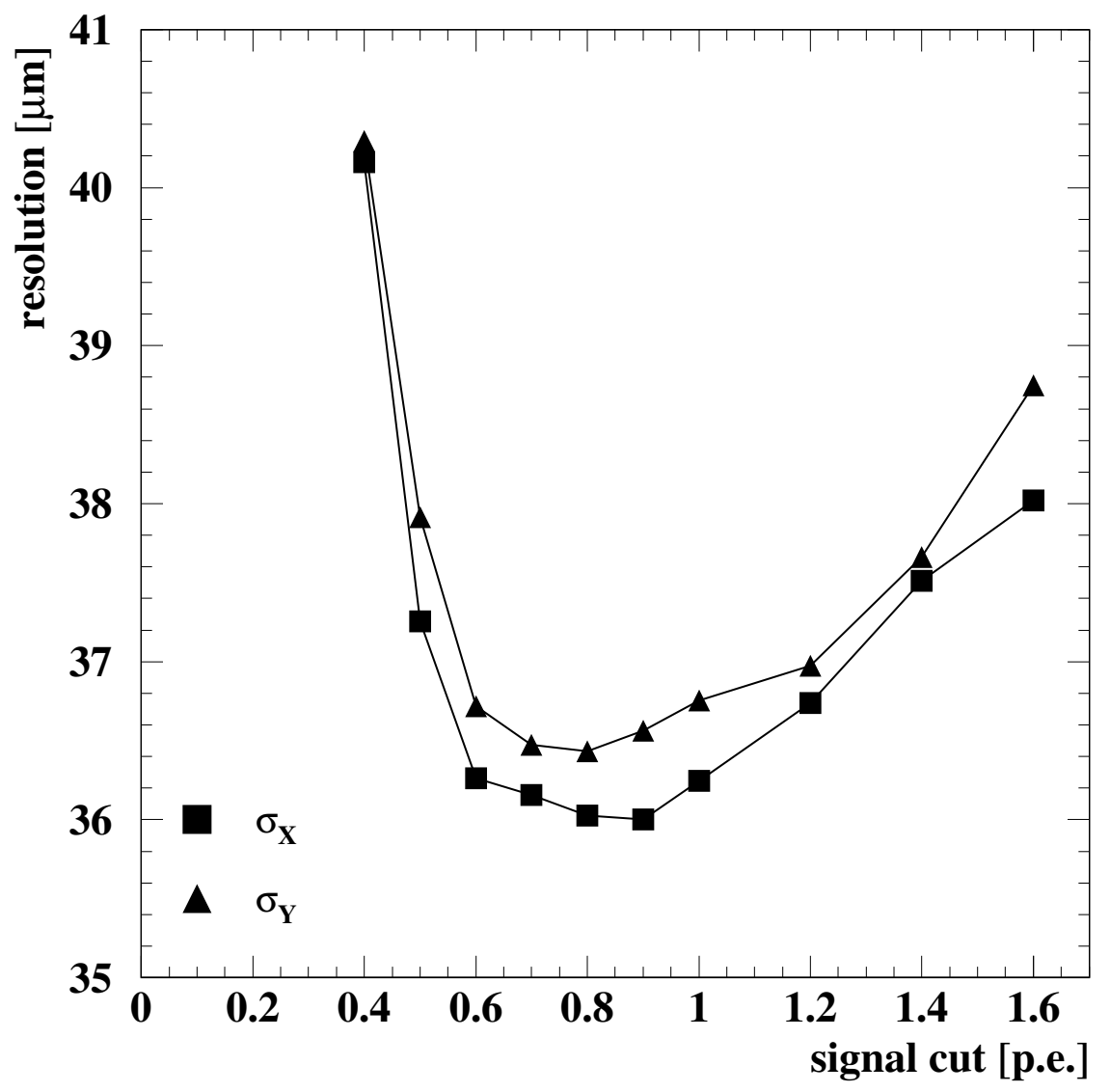

Figure 14. The dependence of the ALFA resolution on the hit-defining cut.

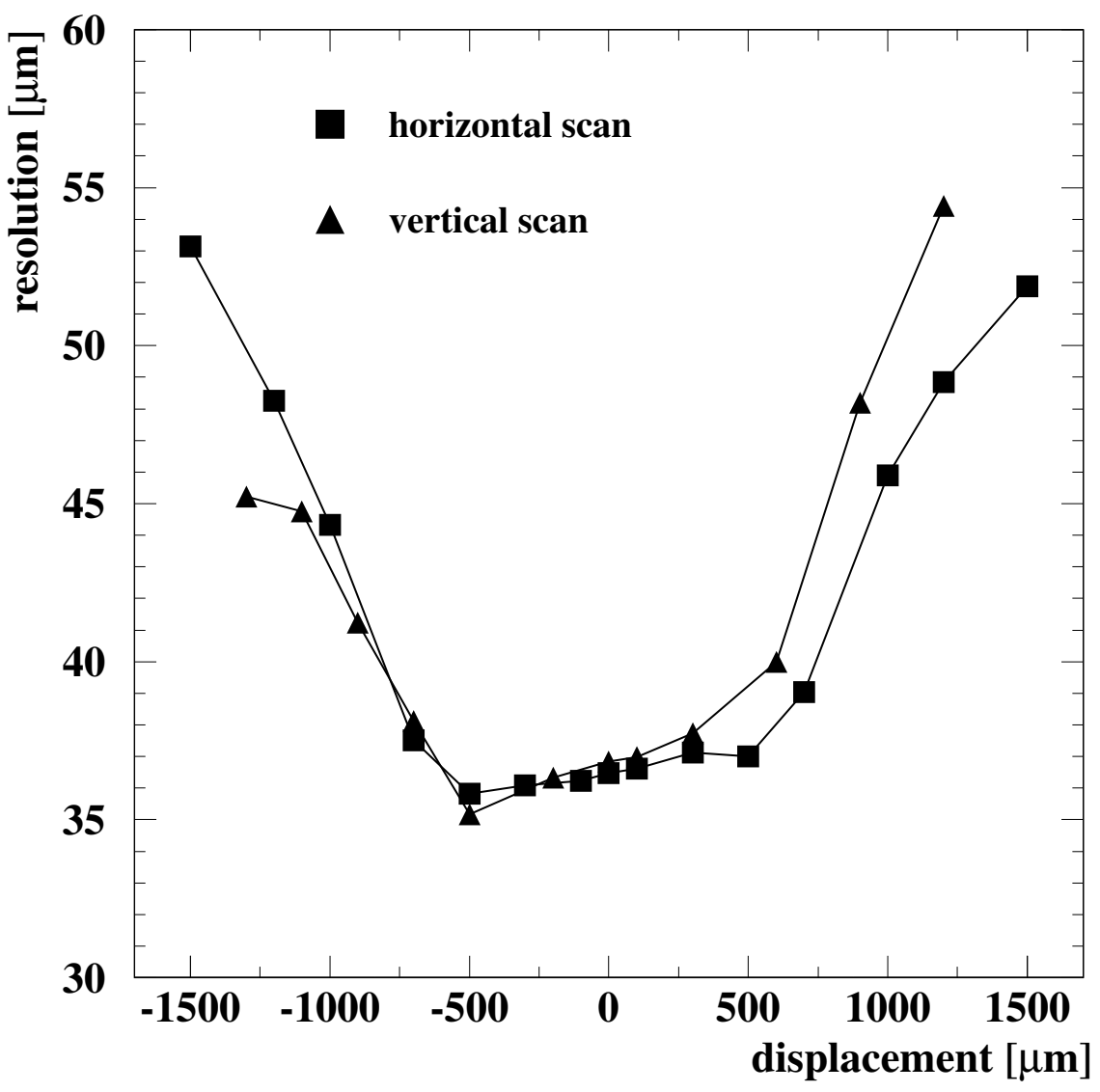

Figure 16. Homogeneity of the ALFA space point resolution for a position scan across the module in vertical and horizontal directions.

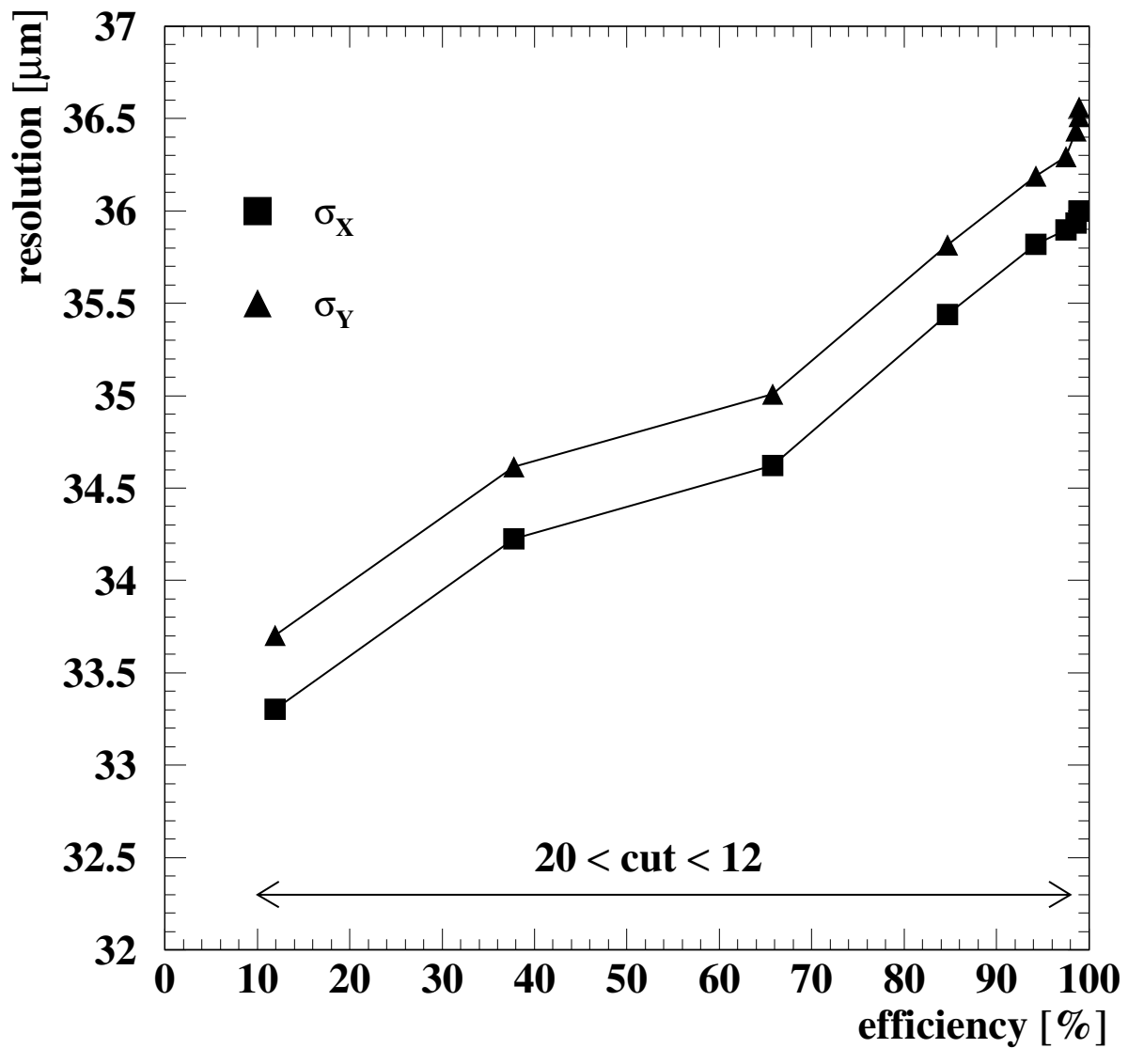

Figure 15. ALFA resolution versus efficiency for an underlying variation of the cut on the minimal number of planes hit, as indicated by the arrow.

The achieved resolution depends only moderatly on the number of hit planes (see Figure 15). For a fixed cut of 0.9 p.e. the resolution improves from about 37 $\mu$m for at least 12 hit planes to about 33.5 $\mu$m when all 20 planes are requested to be hit. The tracking efficiency decreases however drastically from 99% for 12 planes to 10% for 20 planes.

All of the tracking studies carried out above were restricted to the central area of the detector away from the edges. The track resolution degrades at the edges where the number of staggered planes available for the reconstruction decreases. This effect is illustrated in figure 16, where the centre of the beam spot of 1 mm radius has been displaced from the centre of ALFA toward the edges, for the horizontal and vertical axis separately. Given the size of the spot the data are not independent, but the trend is clearly visible. In the central region within $\pm 500$ $\mu$m the performance is very homogeneous. Toward the horizontal edges the resolution degrades steeply, essentially symmetrically along both directions. The vertical scan ending at -1300 $\mu$m covers the cut edge region. The resolution there is only moderately degraded (45 $\mu$m compared to 36 $\mu$m in the centre).

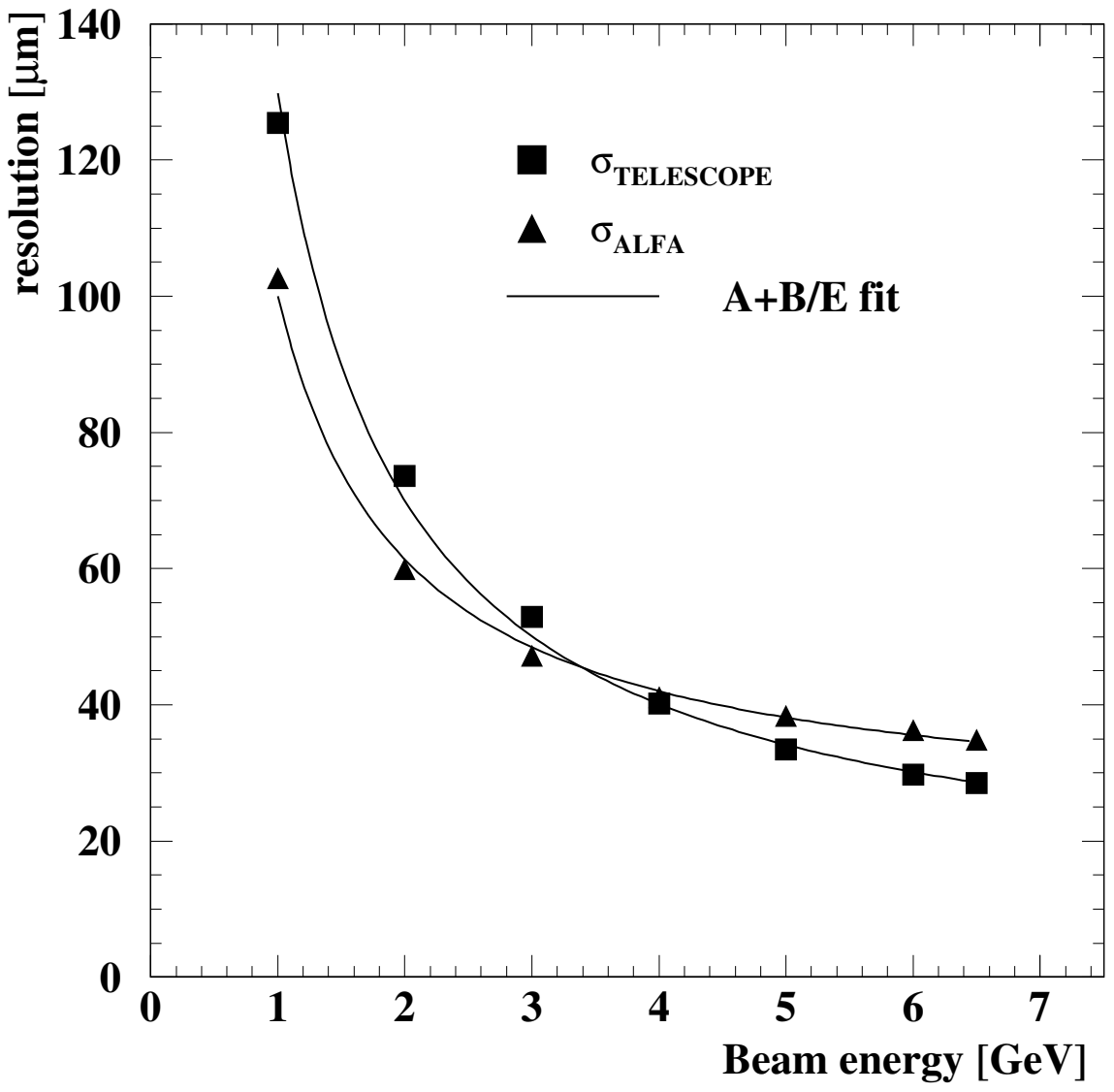

Figure 17. The space point resolution of ALFA and of the telescope as function of the electron beam energy, compared to a fit of a 1/E form.

Finally a scan of the energy of the electron beam in the range from 1 to 6.5 GeV was performed. The relatively low electron energy and the amount of material in the beam line ($\approx 10\% X_0$) generate multiple scattering and shower processes for the beam particles, which are irrelevant for protons at LHC energies. The evolution of the track resolution with the beam energy of both the Si-telescope and ALFA, averaged over the $x$ and $y$ coordinates, is shown in figure 17. A parametric form $A + B/E$, inspired by the energy dependence of multiple scattering, describes well the data. In particular the telescope resolution exhibits a strong energy dependence, as two stations at the end of the beam line are sensitive to multiple scattering occurring upstream e.g. in ALFA. The energy evolution of the ALFA resolution is less steep but indicates that in the LHC regime a tracking resolution of better than 30 $\mu$m is feasible.

### 3.2.5. Edge sensitivity

The edge sensitivity was studied for both the ALFA 2_2_32 and ALFA 10_2_6 detectors with the ALFA edge well inside the standard $S_1 \times S_2 \times S_3$ trigger region. Both detectors were mounted together with an additional scintillator, called S330, touching ALFA from below and which was used to define the physical edge of the fibres. S330 was a scintillator of size $30 \times 3 \times 6$ mm$^3$ $(x, y, z)$. Special runs were taken where the signal from S330 was used in coincidence with $S_1 \times S_2 \times S_3$ as trigger.

For the edge study of ALFA (S330) the $x$ and $y$ information from the Si-telescope was interpolated to the z-position of the ALFA detector. The telescope coordinate system was aligned to coincide with the one of ALFA. Only tracks with polar angles within $\pm 0.5$ mrad around zero were accepted. A 5-$\sigma$ pedestal cut was applied to the ALFA signal amplitudes. As the S330 signal was used directly in the trigger no further selection could be done based on S330 information.

From the tracks passing the selection, histograms were produced of their y-coordinate values given by the Si-telescope. The edge characteristics of the data was then extracted by fitting a *"smeared edge"* function using `MINUIT`. The smeared edge function consists of a Heaviside function convoluted with a Gaussian.

$$f(y; N, c, \sigma) = N \int_{-\infty}^{\infty} H(a') g(a' - a) da' \quad (4)$$

$$= \frac{N}{2} \left( 1 + erf(\frac{a}{\sqrt{2}\sigma}) \right) \quad (5)$$

The fit provides an estimate of the normalization constant $N$, the Heaviside edge location $c$, with $a = c + y$, and the Gaussian smearing $\sigma$.

Both ALFA detectors used in this test show a technical imperfection which will be avoided during the construction of the final detectors. The fibres in one plane were all machined to the same length, however the length of the different planes varied by up to $\pm 75 \mu$m. Therefore the scintillator S330 was always touching only one plane of an ALFA detector.

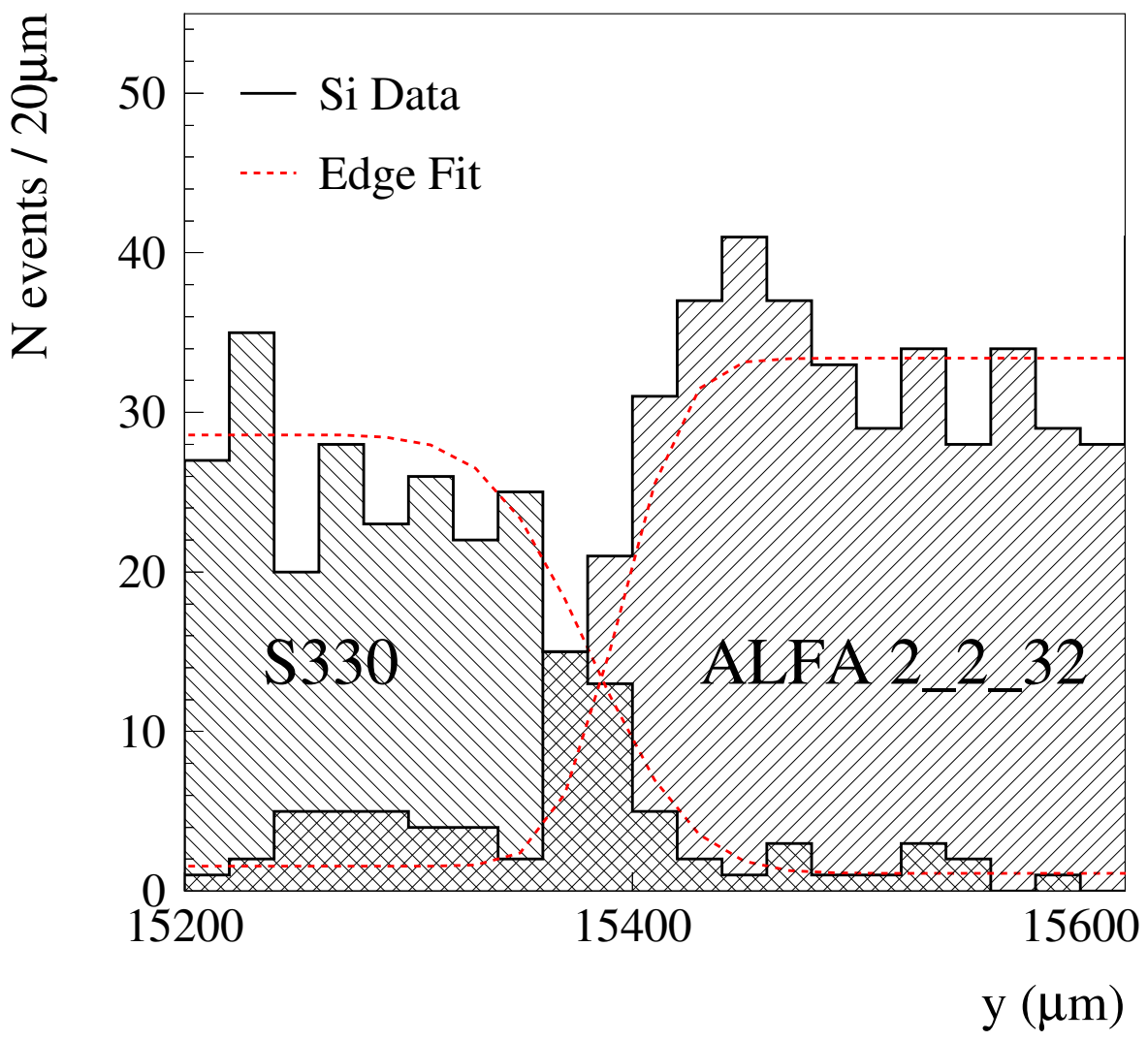

Figure 18. Comparison between the ALFA 2_2_32 and S330 edges. The histograms corresponds to the $y$-coordinate distribution of tracks measured by the Si-telescope. The dashed lines represent the smeared edge functions fitted to the histograms.

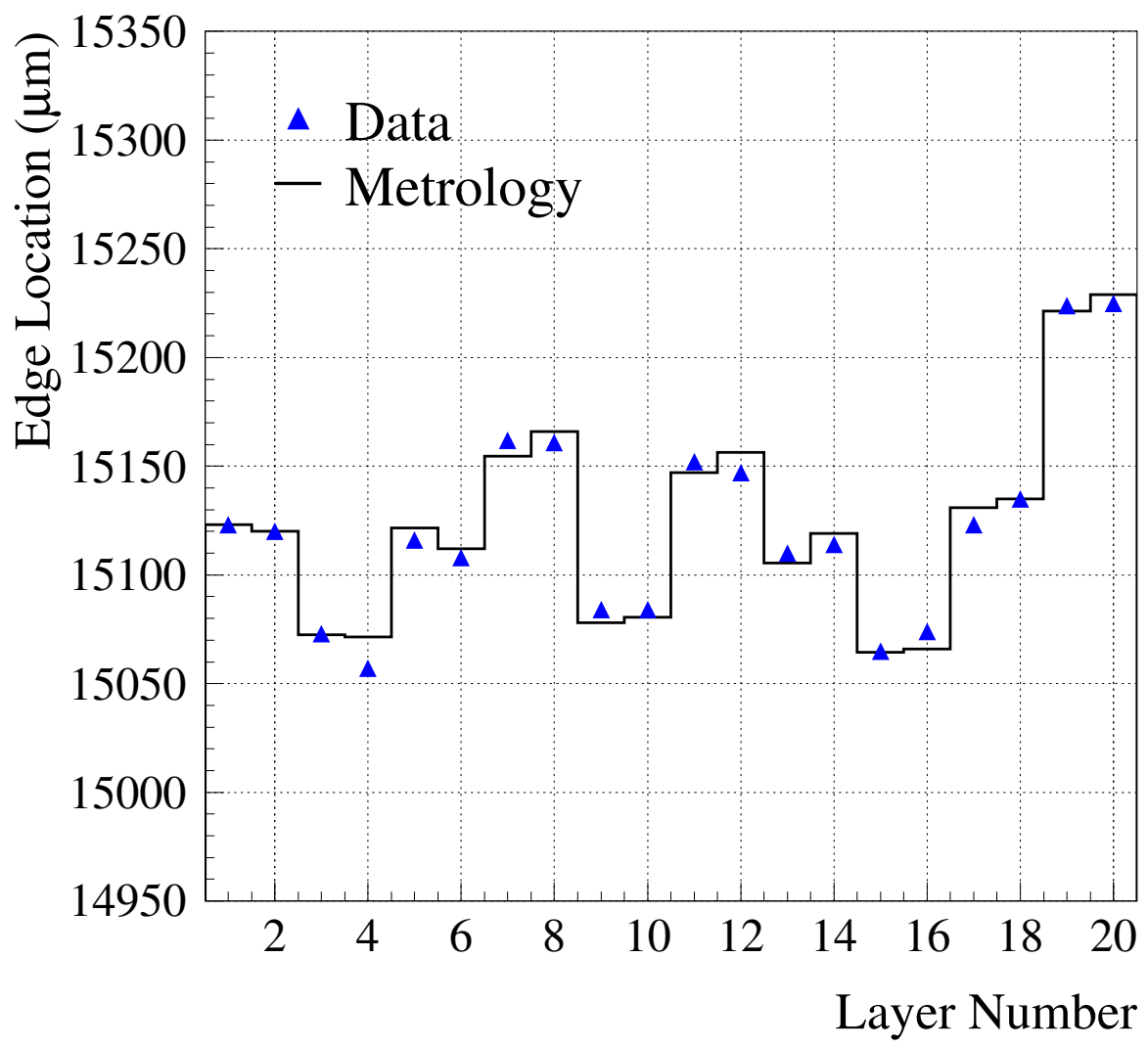

Figure 19. Comparison between the edge locations of ALFA 10_2_6 obtained by the fit and metrology.

Figure 18 shows the $y$-coordinate histograms from the tracks giving a hit in layers 3 or 4 of ALFA 2_2_32 and those traversing S330. The fitted edge functions are also shown. The fit determines the edge locations of layers 3 and 4 at $y = 15386$ and $15393$ $\mu$m, respectively (stat. error $\pm 4$ $\mu$m), and the edge location of S330 at $15371 \pm 6$ $\mu$m. The inactive edge zone at the end of the fibre is much smaller than 100 $\mu$m, it is practically almost compatible with zero. The weighted mean of the Gaussian smearing of the 4 edge functions fitted to ALFA 2_2_32 was found to be $27 \pm 4$ $\mu m$ which is consistent with the Si-telescope resolution of 22 $\mu m$ (for this specific experimental condition).

The precision and reliability of the above method is illustrated in figure 19 which shows the fitted edge locations of the 20 fiber layers of the of the ALFA 10_2_6 detector. The solid line represents the result of a metrological analysis of the individual fiber planes performed with an optical coordinate measurement machine before the beam test. The agreement of the two data sets is as good as 6 $\mu$m (RMS). The weighted mean of the $\sigma$ values from the 20 fits of ALFA 10_2_6 corresponds to $36 \pm 1\mu$m which, compared to a Si-telescope resolution of 29 $\mu$m, also proves the absence of any significant low efficiency region at the fibre edges.

## 4. Conclusions

ATLAS intends to measure the LHC luminosity using elastic scattering at very small angles. For this purpose, we have developed and constructed several prototype detectors to test a detector concept for such measurements. We have chosen scintillating fibre technology as a baseline. This is a simple and cost effective technology and a scintillating fibre tracker fulfills the main requirement of being edgeless and insensitive to signal pick up from the beam. The prototype detectors were extensively tested in a 6 GeV electron beam at DESY. The test clearly validated the detector concept, the construction method and the choice of the single cladded 0.5 mm square baseline fibre

Kuraray SCSF-78. The applied fibre machining and coating techniques were found to be adequate for the required surface quality and reflectivity as was the geometrical precision obtained with the relatively simple tooling. Agreement to better than 5% was found between the calculated light yield and the actual number of photoelectrons measured in the test beam. The optical cross talk was small (1.3%) and appeared to have negligible impact on the detector preformance. Single fiber efficiencies were found to be well above 90%. A reconstruction algorithm was developed that gave tracking efficiencies above 99% and a space resolution of 36 $\mu$m. Moreover, the efficiencies at the edges of the fibres were studied in detail and no significant drop of efficiency was found within the resolution of the method (order 30 $\mu$m). All these parameters match well the ATLAS requirements. We conclude that the proposed technique is fully appropriate for the purpose of the luminosity measurement.

We are now going ahead with the final design and we will construct a full scale fibre plane prototype this year. A test beam run in a high energy proton beam at CERN is being prepared for the autumn 2006. We hope to be able to integrate the new prototype in a Roman Pot unit and to demonstrate the spatial resolution in a high energy proton beam. We also intend to use electronics and DAQ components that are closer to the final application.

**Acknowledgements**

We would like to thank our technical personnel at CERN for excellent work during the development and construction of the fibre tracker: C. David, A. Folley, L. Kottelat, J. Mulon and M. van Stenis. The analytical and finite element calculations by P. Wertelaers (CERN) were very helpful for the validation of the detector design. We appreciated the accurate work of the CERN metrology team led by A. Cherif. The hospitality of DESY and the professional attitude of the accelerator operators are greatly acknowledged. Special thanks to N. Meyners and U. Kötz (both DESY) for their excellent support during the test beam period.